\pdfoutput=1

\documentclass[preprint]{ptephy}

\preprintnumber{} 

\usepackage{boites}

\usepackage{amsmath} 
\usepackage{amsthm} 
\usepackage{graphics} 
\usepackage{algorithmic} 
\usepackage{subfig} 
\usepackage{url} 
\usepackage{footnote}
\usepackage{longtable}
\usepackage{bm}
\usepackage{dcolumn}
\usepackage{hyperref}

\begin{document}
\title{Measurement of the muon beam direction
and muon flux for the T2K neutrino experiment}

\author{\name{K.~Suzuki}{1,\ast}, 
\name{S.~Aoki}{2}, 
\name{A.~Ariga}{3}, 
\name{T.~Ariga}{3},
\name{F.~Bay}{3,}\thanks{Present address: Institute for Particle Physics, ETH Zurich, Zurich, Switzerland}
\name{C.~Bronner}{4},
\name{A.~Ereditato}{3},
\name{M.~Friend}{5}, 
\name{M.~Hartz}{4,8}, 
\name{T.~Hiraki}{1}, 
\name{A.K.~Ichikawa}{1}, 
\name{T.~Ishida}{5},
\name{T.~Ishii}{5},
\name{F.~Juget}{3,}\thanks{Present address: Institute of Radiation Physics, University Hospital and University of Lausanne, Lausanne, Switzerland},
\name{T.~Kikawa}{1,}\thanks{Present address: RCNP, Osaka University, Ibaraki, Osaka, Japan}, 
\name{T.~Kobayashi}{5}, 
\name{H.~Kubo}{1}, 
\name{K.~Matsuoka}{1,}\thanks{Present address: KMI, Nagoya University, Nagoya, Japan},
\name{T.~Maruyama}{5}, 
\name{A.~Minamino}{1}, 
\name{A.~Murakami}{1,}\thanks{Present address: Toshiba Corporation, Kawasaki, Japan}, 
\name{T.~Nakadaira}{5},
\name{T.~Nakaya}{1},
\name{K.~Nakayoshi}{5},
\name{Y.~Oyama}{5},
\name{C.~Pistillo}{3},
\name{K.~Sakashita}{5}, 
\name{T.~Sekiguchi}{5}, 
\name{S.Y.~Suzuki}{5}, 
\name{S.~Tada}{5}, 
\name{Y.~Yamada}{5},
\name{K.~Yamamoto}{6}, 
and \name{M.~Yokoyama}{7}}

\address{\affil{1}{Department of Physics, Kyoto University, Kyoto, Japan}
\affil{2}{Kobe University, Kobe, Japan}
\affil{3}{University of Bern, Albert Einstein Center for Fundamental Physics, Laboratory for High Energy Physics (LHEP), Bern, Switzerland}
\affil{4}{Kavli Institute for the Physics and Mathematics of the Universe (WPI), Todai Institutes for Advanced Study, University of Tokyo, Kashiwa, Japan}
\affil{5}{High Energy Accerlator Research Organization (KEK), Tsukuba, Ibaraki, Japan}
\affil{6}{Department of Physics, Osaka City University, Osaka, Japan}
\affil{7}{Department of Physics, University of Tokyo, Tokyo, Japan}
\affil{8}{TRIUMF, Vancouver, British Columbia, Canada}
\email{k.suzuki@scphys.kyoto-u.ac.jp}}


\begin{abstract}%
The Tokai-to-Kamioka (T2K) neutrino experiment measures neutrino oscillations
by using an almost pure muon neutrino beam produced at the J-PARC accelerator facility.
The T2K muon monitor was installed to measure the direction and stability 
of the muon beam which is produced together with the muon neutrino beam.
The systematic error in the muon beam direction measurement 
was estimated, using data and MC simulation,
to be 0.28~mrad.
During beam operation, 
the proton beam has been controlled 
using measurements from the muon monitor
and the direction of the neutrino beam has been tuned
to within 0.3~mrad 
with respect to the designed beam-axis.
In order to understand the muon beam properties,
measurement of the absolute muon yield at the muon monitor 
was
conducted with an emulsion detector.
The number of muon tracks was measured to be 
$(4.06\pm0.05)\times10^4$~cm$^{-2}$ normalized with $4\times10^{11}$~protons on target
with 250~kA horn operation.
The result 
is in agreement with the prediction which is corrected based on 
hadron production data.
\end{abstract}


\maketitle


\section{Introduction\label{sec:intro}}

The Tokai-to-Kamioka (T2K) experiment \cite{t2k} is a long baseline neutrino oscillation experiment in Japan.
The neutrino oscillation parameters are determined by measuring 
an 
accelerator-produced
neutrino beam before oscillation with the near detector
and near the oscillation maximum with the far detector.
T2K began operation in January 2010.
Since then, data corresponding to a total of $6.63\times10^{20}$~protons on target (p.o.t.)
 had been collected up to May 2013.

The T2K muon monitor \cite{mumon} 
was installed to monitor 
the muon beam
which is produced together with 
the neutrino beam from the decay of pions.
As the muon monitor is the only detector 
to monitor the beam spill-by-spill,
our strategy is to monitor the muon beam direction with a precision of 0.3~mrad
for every beam spill,
in order to better control the neutrino beam for
 the neutrino oscillation measurement.

In this paper, we first provide an overview of the T2K experiment
and the importance of a precise measurement of the muon beam direction
in Sec.~\ref{sec:t2kexperiment}.
Section~\ref{sec:detector} gives
an overview of the components of the muon monitor. 
A method 
for reconstructing the profile of the muon beam with the muon monitor
is described in Sec.~\ref{sec:measure}.
In this section we also show the systematic error in the beam direction measurement,
which was estimated using both the actual beam data and MC simulation.
The stability of the beam direction and its intensity during the T2K beam operation is 
discussed in Sec.~\ref{sec:operation}.
During the beam operation, measurements of the absolute muon yield were conducted 
using an emulsion detector.
This result, and a comparison with the MC prediction, are shown in Secs.~\ref{sec:emulsion}
and \ref{sec:mc} respectively.

\section{Overview of the T2K experiment\label{sec:t2kexperiment}}

\begin{figure}[tb]
\centering
\includegraphics[width=1.0\textwidth]{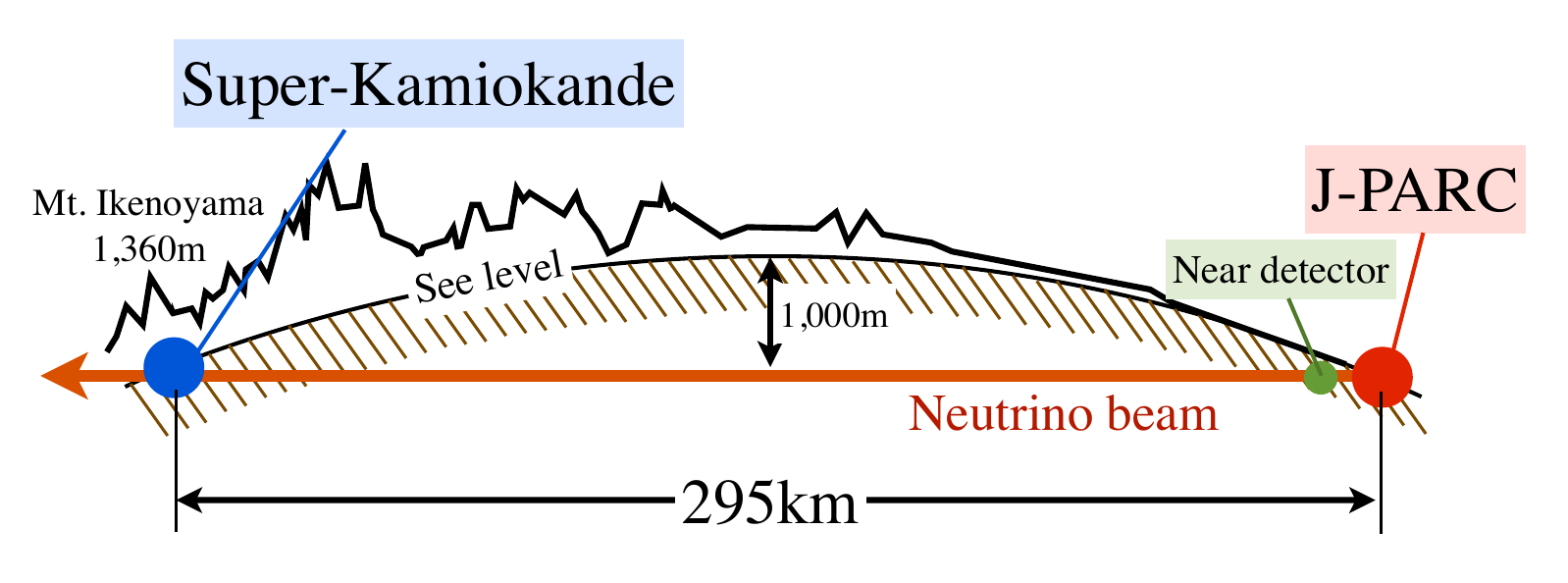}
\caption{Overview of the T2K experiment.}
\label{fig:t2k_xsec}
\end{figure}

\begin{figure}[tb]
\centering
\includegraphics[width=0.9\textwidth]{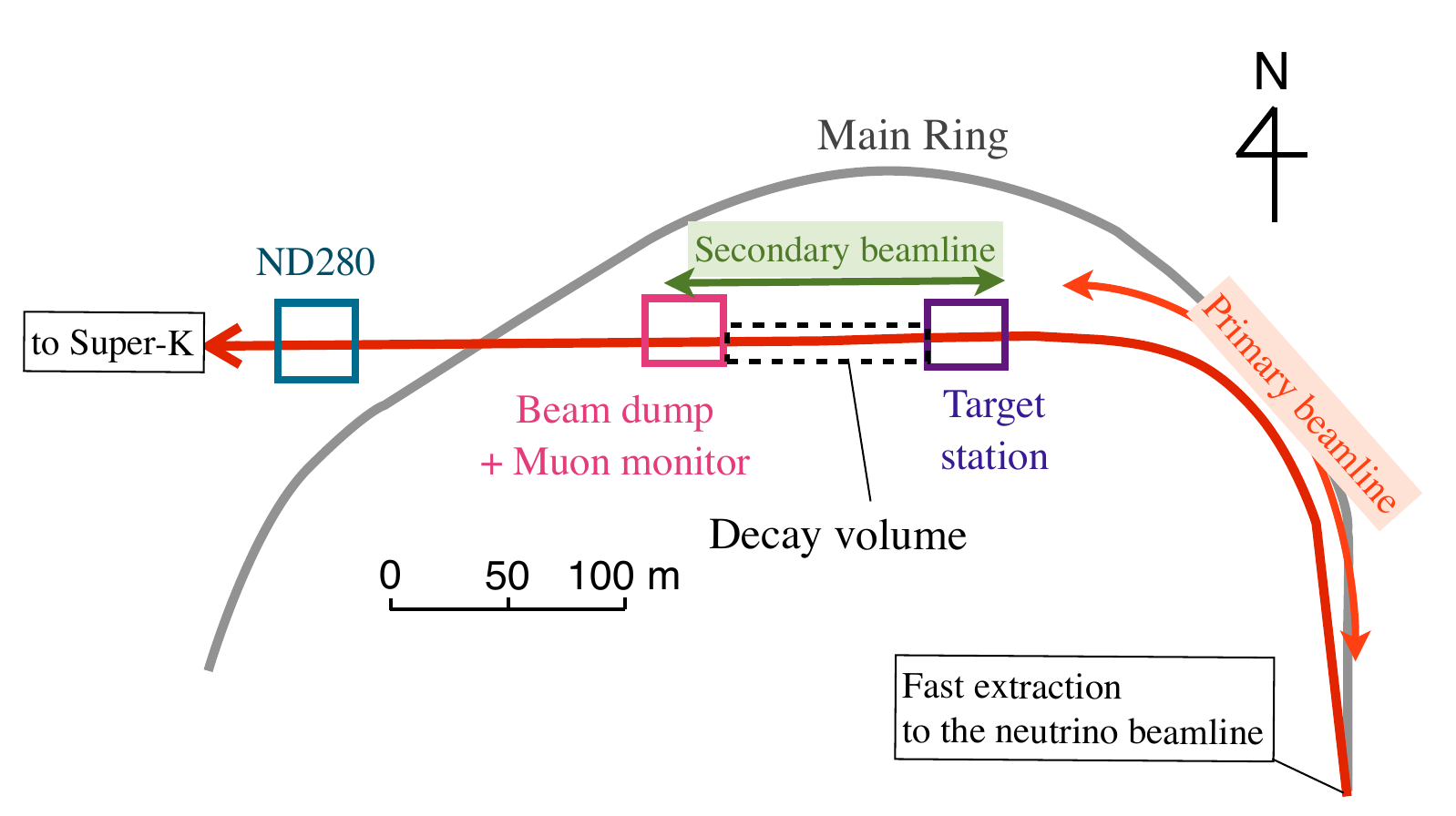}
\caption{Overview of the T2K beamline.}
\label{fig:beamline}
\end{figure}

T2K consists of:
a neutrino beamline,
producing an intense muon neutrino beam;
a near detector complex, INGRID and ND280;
and a far detector, Super-Kamiokande (Super-K).
Using this setup, the experiment aims to measure the neutrino oscillation parameters.
An overview of the T2K experiment is shown in Fig.~\ref{fig:t2k_xsec}.
The Japan Proton Accelerator Research Complex (J-PARC)
is a 
facility situated in Tokai, Japan.
A proton beam is accelerated up to 30~GeV by the main ring synchrotron
and is fast-extracted to the neutrino beamline.
The neutrino beamline consists of two components as shown in Fig.~\ref{fig:beamline}:
a primary and secondary beamline.
In the primary beamline,
the proton beam is transported to 
a graphite target
every 2 to 3 seconds.
The beam has a time structure of eight narrow bunches,
58~ns long with 581~ns intervals,
 in a single spill.
The beam forms a 
two dimensional
Gaussian distribution 
of $\sim$4~mm 
1$\sigma$
width corresponding to $\sim$7~mm radius at the target.
The target and other equipment 
used to produce the
neutrino beam is situated 
in the secondary beam line, whose details are given in Sec.~\ref{subsec:nubeam}.
The neutrino beam produced here
is detected at ND280 and Super-K,
and the oscillation parameters are then measured.

\subsection{Creation of the neutrino beam at the secondary beamline\label{subsec:nubeam}}

\begin{figure}[tb]
\centering
\includegraphics[width=0.9\textwidth]{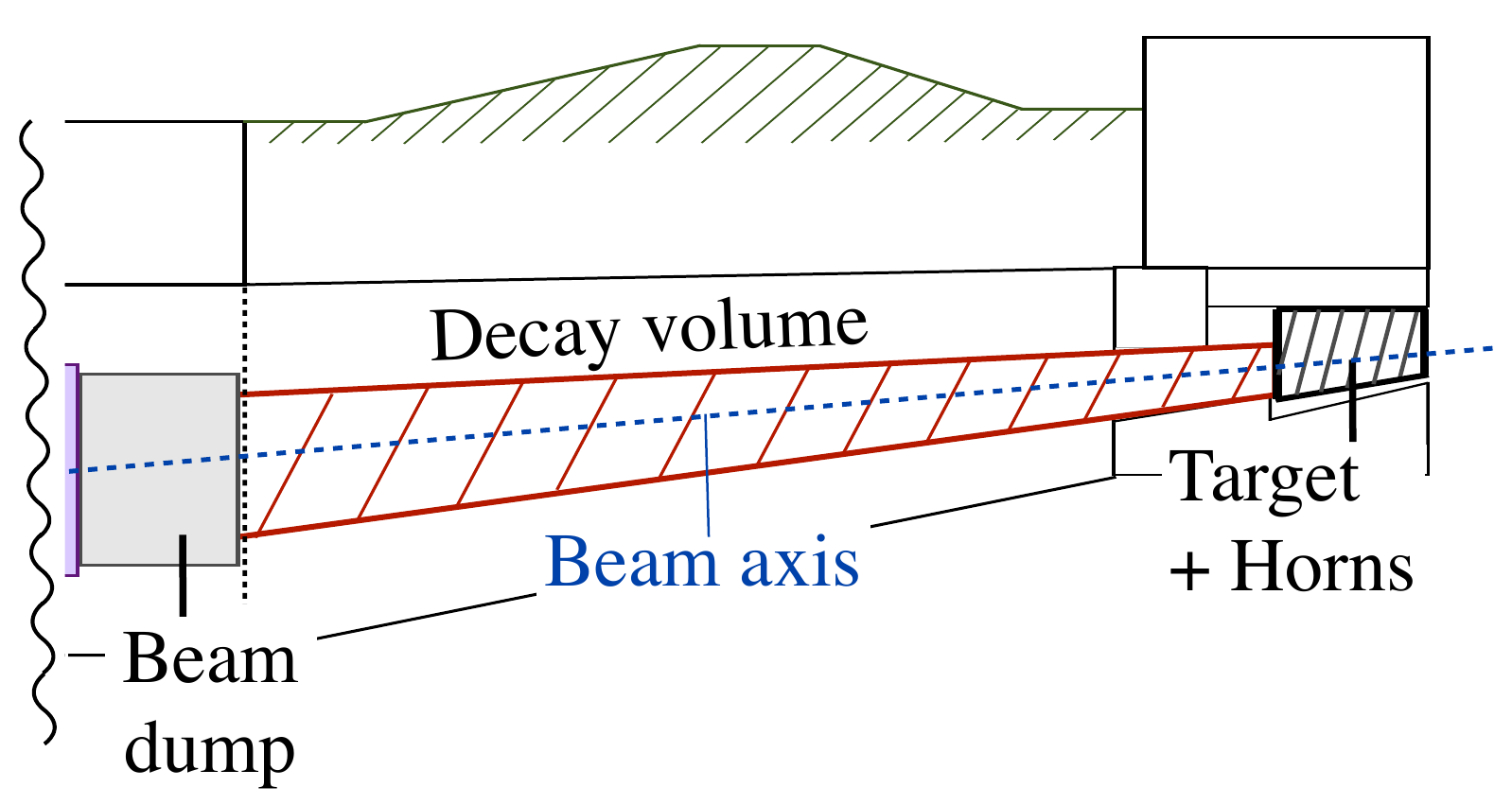}
\caption{Overview of the secondary beamline.
All of the components in the beamline, the target, horns, 
decay volume and beam dump, are contained in a single volume of 
$1500$~m$^3$ filled with helium gas.}
\label{fig:secondary_beamline}
\end{figure}

Figure~\ref{fig:secondary_beamline} provides
an overview of the secondary beamline.
All of the components in the beamline
are contained in a single volume of $\sim1500$~m$^3$ filled with helium gas,
which is enclosed by a helium vessel.
The proton beam,
 transported to the target via the primary beamline,
first enters a baffle which works as a collimator.
After 
passing through the baffle, the proton beam hits the target 
and produces secondary particles, mostly pions.
Three magnetic horns~\cite{HORN}
are used to focus (defocus) these positively (negatively) charged pions
along the designed beam-axis.
Each of the horns is made of aluminum conductor 
and produces a maximum toroidal magnetic field of 1.7~T
inside the conductor when the horns operate at 250~kA.
The decay volume for the pions is 
a 96~m long steel tunnel.
A pure and intense muon neutrino beam is produced as the pions decay in 
this tunnel.
The beam dump sits at the end of the decay volume
to absorb the hadron flux from the beam.
It consists of a core 
composed
of 75~tons of  graphite and 
fifteen (two) iron plates placed outside (inside) the helium vessel at the downstream end of the core.
The length is 3.174~m, in line with the beam-axis.

\subsection{Importance of measuring the neutrino beam direction and intensity}

Both ND280 and Super-K are located 2.5~degrees
 from the 
beam-axis.
This experimental setup enables us to utilize a narrow-band neutrino beam 
with a peak energy
around 0.6~GeV at which neutrinos oscillate with near the maximum probability
after traveling 295~km.
However, a 1~mrad uncertainty in the beam direction measurement 
leads to a 2-3~\% uncertainty in the neutrino energy scale,
affecting measurements of the neutrino oscillation parameters.
Therefore the beam direction has to be monitored with a good precision
 and controlled well to eliminate the additional uncertainty 
 for the oscillation parameters.
 In addition, a contingency may arise in the beamline during operation,
 such as a sudden drop in the horn currents or deterioration of the target,
resulting in a decrease in the neutrino beam intensity.
 Therefore, monitoring not only the direction but also 
 the intensity of the neutrino beam has to be 
 done 
on a spill-by-spill basis in order to 
  promptly confirm the state and health
of the beamline components as well as quality of the neutrino beam.

 \subsection{Beam monitors}

T2K employs two beam monitors for the beam direction measurement,
INGRID and the muon monitor.
INGRID~\cite{ingrid} is located 280~m downstream of the target.
It has 14 independent modules which are composed of 
sandwiched iron plates 
and scintillator planes. 
The modules are installed at positions in cross shape 
centered on the beam-axis.
The profile of the neutrino beam is reconstructed by 
counting the number of neutrino interactions 
in each of the modules.
Due to the small cross section of the neutrino interactions,
the time for accumulating neutrino events depends on the beam intensity 
and typically requires one day for the profile reconstruction with 
a proton beam power of $\sim$100~kW.

The T2K muon monitor is another beam monitor
that
monitors the muon beam 
which is produced together with the neutrino beam from the decay of pions.
The monitor is located 118~m downstream of the target 
and just downstream of the beam dump which absorbs the hadron flux.
Unlike the INGRID measurement, 
the muon monitor can detect the muon beam spill-by-spill.
Therefore, the intensity and direction of neutrino beam can be 
indirectly
monitored with the muon monitor spill-by-spill.
It is 
necessary to monitor the muon beam direction with a precision of 0.3~mrad
in order to control the neutrino beam direction to 
within 0.3~mrad with respect to the beam-axis.


%
%
%

\section{Instrumentation of the muon monitor \label{sec:detector}}
\begin{figure}[tb]
\centering
\includegraphics[width=0.50\textwidth]{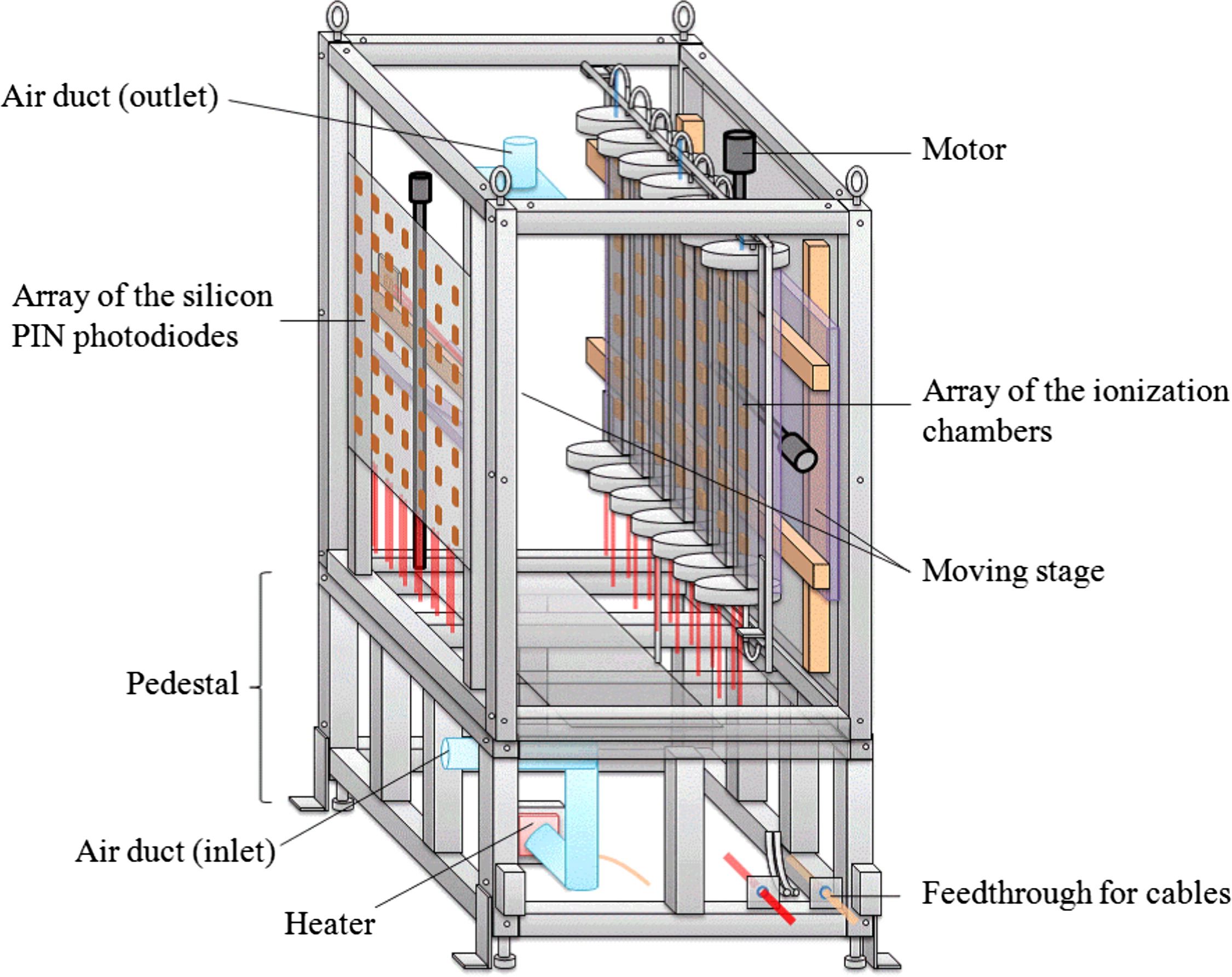}
\includegraphics[width=0.48\textwidth]{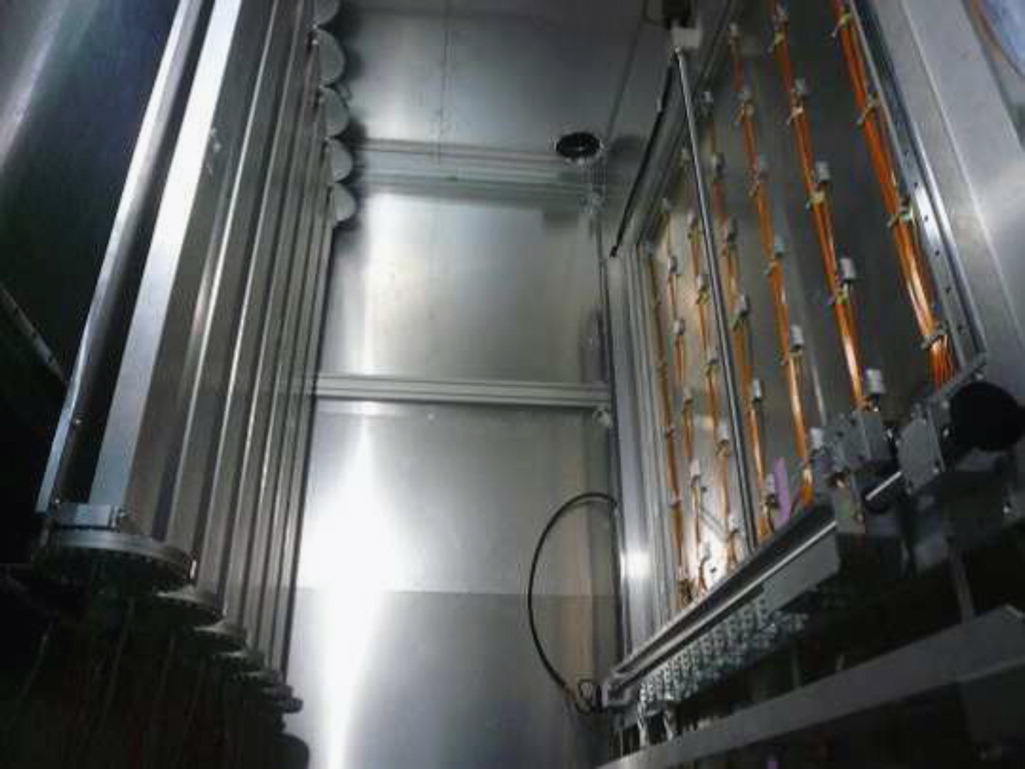}
\caption{Schematic view (left) and the photograph of the muon monitor (right). 
The monitor consists of array of the ionization chamber and the silicon PIN photodiode.
The muon beam enters into the array of the silicon PIN photodiodes first 
and then goes through the array of the ionization chamber~\cite{mumon}.}
\label{fig:mumon}
\end{figure}

The T2K muon monitor was installed in the muon pit 
located just downstream of the beam dump and 18.5~m underground.
Details of the design of the monitor are described in~\cite{mumon}.
The thickness of
the beam dump is chosen 
to minimize the hadron flux while 
retaining the sensitivity 
in the measurement of the muon beam direction;
only muons with energy above 5~GeV 
can pass through the beam dump and reach the muon monitor.
Figure~\ref{fig:mumon} shows a schematic view (left) and photograph (right)
of the muon monitor.
The monitor consists of two independent detectors:
an array of ionization chambers and 
another array of silicon PIN photodiodes.
Each detector array has $7 \times 7=49$ sensors at 25~cm intervals
and covers an area of 150$\times$150~cm$^2$ with respect to the beam-axis.

\subsection{Ionization chamber\label{subsec:IC}}
Each of the seven ionization chambers contains
seven sets of two parallel $100\times100$~mm$^2$
ceramic plates separated by 3~mm. 
One of the two parallel plates has a signal electrode which has
a dimension of
$75\times75$~mm$^2$ and
 is surrounded by the ground electrodes.
A bias voltage of 200~V is applied to
a $93\times93$~mm$^2$ electrode on the other plate
and a uniform electric field is created 
through a $75\times75$~mm$^2$ area
between the two electrodes.
Thus, ionization pairs generated only in the $75\times75\times3$~mm$^3$
 volume contribute to the signal.
 All of the chambers are filled with a gas mixture set to be 
 98\% Ar and 2\% N$_2$ for a beam intensity below $2.3\times10^{13}$~protons per bunch (p.p.b.).
For higher beam intensity,  99\% He and 1\% N$_2$ is used instead as
the size of the signal with He gas is $\sim$10~times smaller than that with Ar gas. 
In both of the gas systems,  N$_2$ is used as a quencher and 
to render the signal insensitive to the amount of impurities in the gas
via the Jesse effect~\cite{jesse}.

\subsection{Silicon PIN photodiode\label{subsec:Si}}
The silicon PIN photodiode 
(HAMAMATSU$^{\textregistered}$ S3590-08)
has an active area of $10\times10$~mm$^2$ and
 a depletion layer thickness of 300~$\mu$m. 
 The silicon layer is mounted on a ceramic base. 
 In order to
 fully deplete the layer, a bias voltage of 80~V is applied.
 The photodiode is put on a PEEK$^{\text{TM}}$ base fixed 
 to the support enclosure and is covered by an aluminum base.

The silicon PIN photodiode is not tolerant of the severe radiation in the muon pit.
There is a report~\cite{Ziock} that the depletion voltage of the silicon PIN detector 
falls 50\% at about $0.7\times10^{13}$~protons/cm$^2$
and reaches a minimum at $1.25\times10^{13}$~protons/cm$^2$.
The decrease in the signal was also reported in the beam test
where a 100~MeV electron beam was used~\cite{mumon}.
From these results, it was estimated that the signal starts to decrease
after a one month exposure to the muon beam in the case of 0.75~MW beam operation.

\subsection{Electronics}
Both signals from the ionization chamber and silicon PIN photodiodes are 
transmitted
by about 70~m of co-axial cables
which connect
the muon pit underground to
an electronics hut on the surface.
The signals are digitized by Flash-ADC modules (FADC) of the COPPER system~\cite{Higuchi:2003xs}
developed by KEK.
The resolution and sampling rate of the FADC are 12~bit and 65~MHz, respectively.
For the signal from the ionization chamber, the gain in the FADC is set to 5.
On the other hand, 
unity gain
is set for the signal from the photodiodes
since the size of the signal is about 30 times larger than 
that of the signal from the ionization chambers.
Instead, the signal
from the photodiode
 is attenuated 
by 0, 15, and 30~dB 
depending on the beam intensity.
Both FADC and signal cable are well calibrated with 
a CAMAC charge/time generator (Phillips 7120) with 1\% precision.

\section{Measurement of the muon beam direction\label{sec:measure}}

The signal from each of the sensors are read out by the FADC and 
integrated to calculate the collected charge.
The profile of the muon beam is then reconstructed by fitting 
the two-dimensional charge distribution with a two-dimensional Gaussian function.
Details of the analysis method are described in Sec.~\ref{subsec:method}.
We also prepare MC simulation 
for the prediction of the muon flux:
simulation of hadronic interactions inside and outside of the target
and calculation of the particle decay
in the decay volume.
Section~\ref{subsec:mc_prep} explains the simulation in more detail.
The sensors
are calibrated using the actual beam 
at the beginning of each beam operation period
as described in Sec.~\ref{subsec:calibration}.
The systematic error for the beam direction measurement was estimated 
using both the actual beam data and MC simulation, this is detailed in Sec.~\ref{subsec:systematics}.

\subsection{Reconstruction of the muon beam profile\label{subsec:method}}

\begin{figure}[tb] 
 \centering
   \includegraphics[width=1.\textwidth]{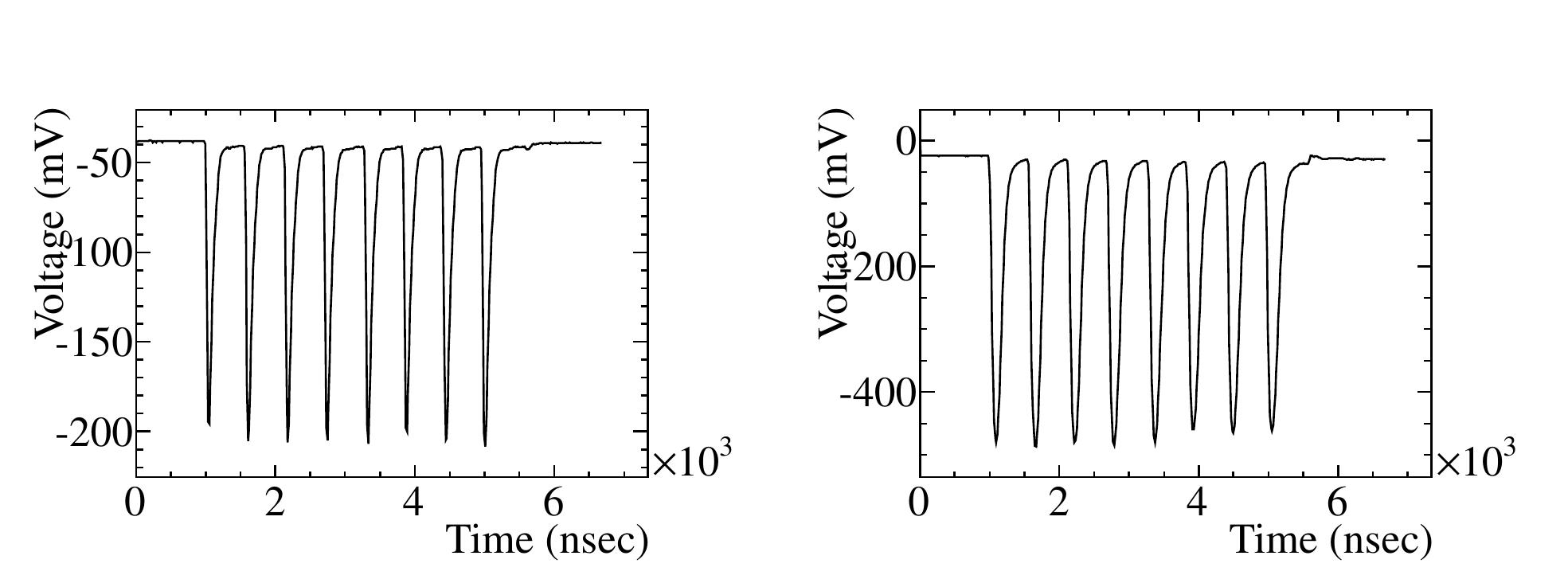} 
   \caption{Waveform of the signal from the silicon PIN photodiode (left)
   and ionization chamber (right), digitized by the FADC.
   Both of the signals are from sensors placed at the center of the arrays and recorded 
   when the beam intensity is $1.3\times10^{13}$~p.p.b.
   The attenuator level was set to 30~dB for the signal from the silicon sensors.}
   \label{fig:waveform}
\end{figure}

\begin{figure}[tb] 
   \centering
   \includegraphics[width=1.\textwidth]{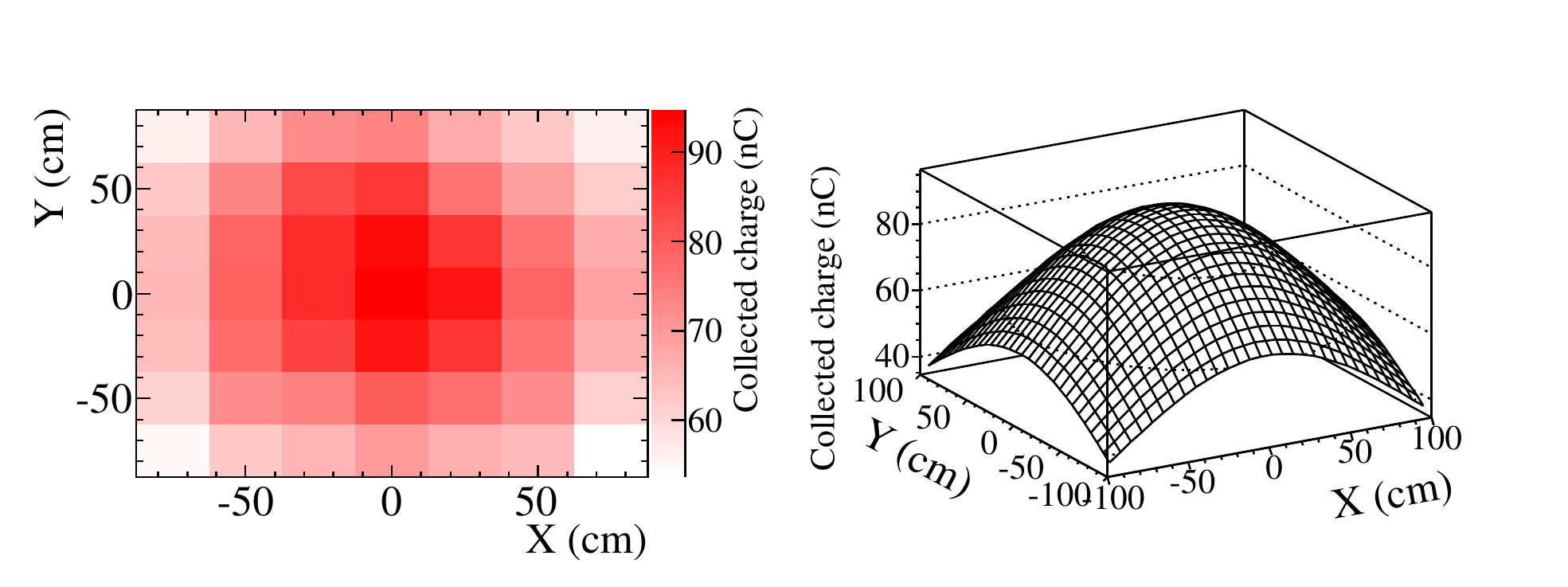}
   \caption{Charge distribution (left) and reconstructed profile (right) of the muon beam measured by the silicon array.
   The collected charge is obtained for each sensor by
   integrating the waveform of all of the bunches (i.e spill) read out by the FADC.
   This is a beam event when the intensity is $1.3\times10^{13}$~p.p.b.}
   \label{fig:siprof}
\end{figure}

\begin{figure}[tb] 
   \centering
   \includegraphics[width=1.\textwidth]{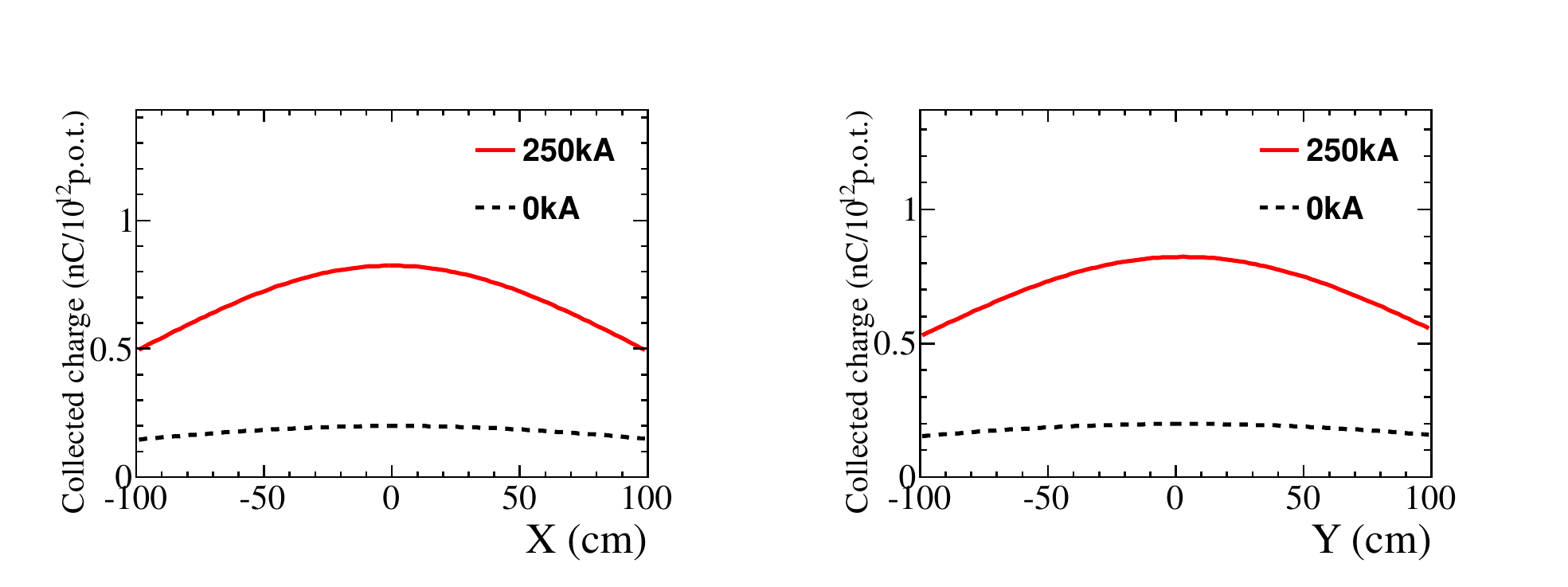}
   \caption{
   Muon beam profile obtained with the silicon array when horns are operated at 0~kA (dashed black)
   and 250~kA (solid red). 
   The horizontal (vertical) profile is shown in left (right).
   }
   \label{fig:siprof_overlay}
\end{figure}

The collected charge is calculated for each sensor by 
integrating a waveform digitized by the FADC.
The typical waveforms recorded during 
beam operation are shown in Fig.~\ref{fig:waveform}.
These waveforms were obtained at a beam intensity of 
$1.3\times10^{13}$~p.p.b. where the attenuator level 
was set to 30~dB for the signal from the silicon sensors.
In the analysis, the integration windows are set to each bunch so that 
the profile of the muon beam can be reconstructed bunch-by-bunch.
Once the collected charge is calculated for each bunch, 
it is then summed over all bunches to get the muon beam direction and intensity.
In this way, the measurement precision statistically improved 
by the square root of the number of bunches.
A left in Fig.~\ref{fig:siprof} shows the charged distribution measured by the silicon PIN photodiodes,
which is obtained by summing the distribution over all bunches.
In order to extract the profile of the muon beam,
the distribution is then fitted by a two-dimensional Gaussian function defined as follows:
\begin{equation}
\label{eq:profilefit}
f(x,y)=A\exp \left[ - \frac{(x-x_0)^2}{2\sigma_x^2} - \frac{(y-y_0)^2}{2\sigma_y^2}\right]
\end{equation}
where $A$ is a normalization parameter;
$x_0$ and $y_0$ represent centers of the beam profile in the horizontal
and vertical direction, respectively;
$\sigma_x$ and $\sigma_y$ represent widths of the beam profile in the horizontal
and vertical direction, respectively.
An example of the reconstructed profile obtained 
from a fit of the two-dimensional Gaussian is shown on the right in Fig.~\ref{fig:siprof}.
Figure~\ref{fig:siprof_overlay} shows 
profiles of the muon beam for the horn currents of 0~kA and 250~kA.
The peak charge 
collected at 250~kA operation is 
about 4 times larger than that 
collected at 0~kA operation.
The muon beam direction $\{\theta_x, \theta_y\}$ is then calculated using parameters $\{x_0, y_0\}$
and distance ($=L$) between the target and muon monitor:
\begin{equation}
\theta_x = x_0/L,  \hspace{2em}
\theta_y=y_0/L \hspace{2em} (L=118\text{~m})
\end{equation}
Here we use an approximation of  
$\tan\theta_{x(y)} \simeq \theta_{x(y)}$,
as $\theta_{x(y)}\ll1$.

\subsection{Monte Carlo simulation\label{subsec:mc_prep}}
The MC simulation consists of two processes:
a simulation of the hadronic interaction in the graphite target,
and propagation of the secondary particles until they interact or decay.
The hadronic interaction in the target is simulated with FLUKA2008~\cite{Ferrari:2005zk}
which was found to be in better
 agreement with external hadron production data
\footnote{Recently FLUKA2011 is found to be the best agreement
with external hadron production data.}$^,$\footnote{The hadron interactions are further tuned with the external experiment data in Sec.~\ref{sec:mc}.}.
Kinematic information for particles emitted from the target is 
saved and transferred to the JNUBEAM simulation~\cite{flux}. 
JNUBEAM is a GEANT3~\cite{GEANT3} 
simulation of 
the secondary beamline including the muon monitor.
The geometry of these components is modeled based on the 
final mechanical drawings of the constructed beamline.
Hadronic interactions are modeled by GCALOR~\cite{GCALOR} in JNUBEAM.
Table~\ref{tab:pid_at_mumon} shows the MC estimation of 
flux of particles
penetrating the muon monitor at the 250~kA horn current setting.
The muon are accompanied by soft components such as gammas and $\delta$-ray electrons.
The particles contributing to the signal at the muon monitor
is also estimated using the MC simulation where argon gas is used for the ionization chamber 
and horn currents are set to 250~kA.
The result is shown in Table~\ref{tab:pid2_at_mumon}.
In both the silicon and ionization chamber arrays,
the muons account for about 80\% of the total amount of the signal.
The subsequent contribution to the signal comes from $\delta$-ray, 
accounting for about 10\% of the total.
A breakdown of the muon flux by the parent particles ($\pi^{\pm}$, $K^{\pm}$ and $K^0_L$)
is shown in Table~\ref{tab:muon_parents}.
As listed in the table,
92\% (95\%) of total $\mu^+$ ($\mu^-$) production is attributable to 
parent $\pi^+$ ($\pi^-$).
Table~\ref{tab:muon_parents_material} shows the breakdown of parent particles 
($\pi^{\pm}$ and $K^{\pm}$)
generated at each of the materials in the secondary beamline.
Most of the pions contributing to the muon flux
are generated at the graphite target.
The subsequent contributions 
from pions 
come from interactions at the beam dump (carbon) which is placed just in front of the muon monitor.

\begin{table}[tb]
\centering
\caption{Breakdown of the particles (/10$^{13}$~p.o.t.) arriving at the muon pit
and going through the area covered by the muon monitor ($150\times150$~cm$^2$).
These are estimated by the MC simulation 
with 250~kA horn current settings.}
\begin{tabular}{lrlrl}
\hline\hline
Particle & \multicolumn{2}{c}{Particles} & \multicolumn{2}{c}{Particles}\\
type  & \multicolumn{2}{c}{at the silicon array} & \multicolumn{2}{c}{at the chamber array }\\
\hline
$\mu^+$  & $2.39\times10^{10}$ & (49.3\%) &  $2.20\times10^{10}$ & (52.0\%) \\
$\mu^-$  & $0.18\times10^{10}$ & (3.7\%) & $0.17\times10^{10}$ & (3.9\%) \\
$e^-$    & $0.32\times10^{10}$ & (6.7\%)  & $0.26\times10^{10}$ & (6.3\%) \\
$e^+$    & $0.03\times10^{10}$ & (0.6\%)& $0.02\times10^{10}$ & (0.6\%) \\
$\gamma$ & $1.92\times10^{10}$ & (39.6\%) & $1.56\times10^{10}$ & (37.0\%)\\
others  & $<0.01\times10^{10}$ & (0.1\%)& $<0.01\times10^{10}$ & (0.2\%)\\
\hline
Total & $4.84\times10^{10}$ & (100\%) & $4.22\times10^{10}$ & (100\%)\\
\hline\hline
\end{tabular}
\label{tab:pid_at_mumon}
\end{table} 

\begin{table}[tb]
\centering
\caption{Breakdown of the particles (/10$^{13}$~p.o.t.) contributing to the 
signal at the muon monitor.
The number listed in the table is estimated for particles going through the area covered by the monitor ($150\times150$~cm$^2$).
In this MC estimation, argon gas is used for the ionization chamber
and horn currents are set to 250~kA.}
\begin{tabular}{lrlrl}
\hline\hline
Particle & \multicolumn{2}{c}{Particles} & \multicolumn{2}{c}{Particles }\\
type  & \multicolumn{2}{c}{at the silicon array} & \multicolumn{2}{c}{at the chamber array }\\
\hline
$\mu^+$  & $2.39\times10^{10}$ & (82.2\%) &  $2.19\times10^{10}$ & (83.4\%) \\
$\mu^-$  & $0.18\times10^{10}$ & (6.1\%) & $0.17\times10^{10}$ & (6.3\%) \\
$e^-$    & $0.30\times10^{10}$ & (10.2\%)  & $0.25\times10^{10}$ & (9.3\%) \\
$e^+$    & $0.03\times10^{10}$ & (0.9\%)& $0.02\times10^{10}$ & (0.9\%) \\
$\gamma$ and others  & $0.02\times10^{10}$ & (0.6\%)& $<0.01\times10^{10}$ & ($<$0.1\%)\\
\hline
Total & $2.90\times10^{10}$ & (100\%) & $2.63\times10^{10}$ & (100\%)\\
\hline\hline
\end{tabular}
\label{tab:pid2_at_mumon}
\end{table}

\begin{table}[tb]
\centering
\caption{Breakdown of the muon flux by the parent particles ($\pi^{\pm}$, $K^{\pm}$ and $K^0_L$)
for the 250~kA horn current setting.}
\begin{tabular}{ll}
\hline\hline
Parent particle & $\mu^+$ ($\mu^-$)\\
\hline
$\pi^+$ ($\pi^-$) & 91.73\% (94.71\%) \\
$K^+$  ($K^-$) & 8.26\% (5.13\%)\\
$K^0_L$ & 0.01\% (0.16\%) \\
\hline\hline
\end{tabular}
\label{tab:muon_parents}
\end{table}

\subsection{Detector calibration\label{subsec:calibration}}

\begin{savenotes}
\begin{table}[tb]
\centering
\caption{Breakdown of the muon parent particles generated at each material for the 250~kA horn
setting.
The last column shows the breakdown for the total flux.
A symbol in a parenthesis denotes the main material element.}
\begin{tabular}{llllll}
\hline\hline
Material & $\pi^+$ & $\pi^-$ & $K^+$ & $K^-$ & Total \\
\hline
Graphite target (C)  & 94.00\% & 64.00\% & 89.80\% & 53.94\% & 91.58\%\\
Horn (Al)            & 1.52\%  & 3.92\%   &  1.16\%  & 1.02\% & 1.65\%\\
Decay volume (He)    & 1.24\%  & 6.70\%   &  0.99\%  & 3.34\% & 1.59\%\\
Decay volume (Fe)    & 0.35\%  & 3.94\%   &  0.65\%  & 2.83\% & 0.61\%\\
Beam dump (C)        & 2.36\%  & 20.80\%  &  7.15\%  & 38.36\% & 4.06\%\\
Other materials      & 0.53\%  & 0.63\%   &  0.25\%  & 0.51\% & 0.51\%\\
\hline
& 100\% & 100\% & 100\% & 100\% & 100\%\\
\hline\hline
\end{tabular}
\label{tab:muon_parents_material}
\end{table}
\end{savenotes}

Both of the detectors, silicon PIN photodiodes and ionization chambers,
are relatively calibrated using the real beam 
at the beginning of each beam operation period.
The ionization chamber is calibrated by moving the entire chamber array 
by $\pm$25~cm 
in both the horizontal and vertical direction
and measuring the muon profile at nine different configurations\footnote{
The 25~cm is equal to the interval of the sensors.}.
These nine measured profiles should be the same on the assumption 
that the muon beam profile itself does not change
over the course of the measurements.
In this way the 49 sensors are relatively calibrated with a precision of 0.4\%.

\begin{figure}[tb] 
   \centering
   \includegraphics[width=0.5\textwidth]{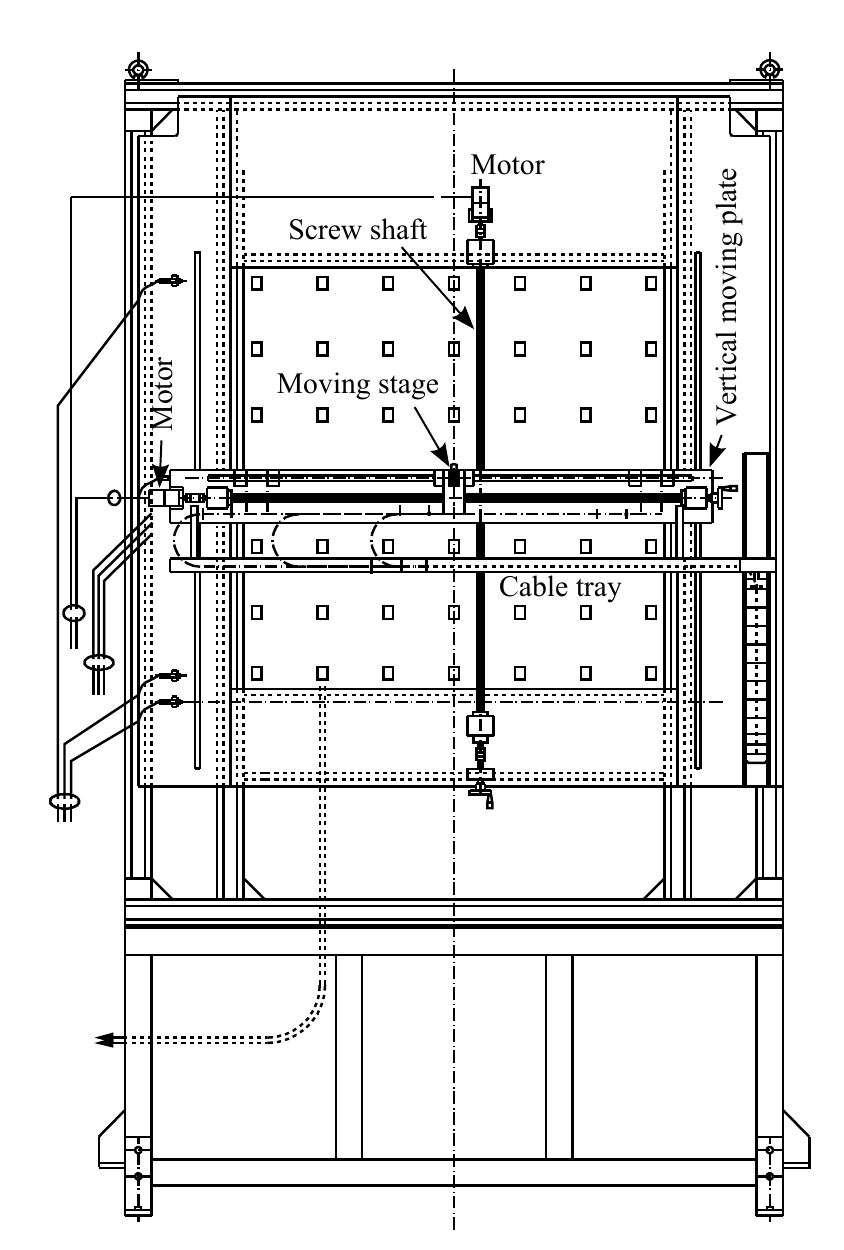} 
   \caption{Moving stage for the calibration silicon PIN photodiode, as viewed from downstream.}
   \label{fig:si_move_stand}
\end{figure}

The silicon PIN photodiodes are calibrated sensor-by-sensor
by measuring the muon beam with an extra 
calibration sensor mounted on small moving stage 
behind the silicon array (see Fig.~\ref{fig:si_move_stand}).
The calibration sensor is placed 
behind each sensor and
the collected charge ratio of the moving sensor to each of the other sensors
is measured:
\begin{equation}
R_i = Q_i/Q_{ref} \hspace{1em}(i=1,2,..,49)
\end{equation}
where $Q_{ref}$ and $Q_{i}$ are 
the collected charges obtained by the extra calibration sensor and $i^{th}$ signal sensor, respectively.
The correction factor is then calculated for each sensor using the mean of the charge ratios:
\begin{equation}
G_i = \langle R \rangle/R_i
\end{equation}
This 
correction
is then applied to each sensor.
In this way, all of  the sensors are calibrated with a precision of 0.1\%.

\subsection{Systematic error in the beam direction measurement\label{subsec:systematics}}

The systematic error in the beam direction comes from:
\begin{enumerate}
\item
uncertainty of the structure of the upstream materials,
\item
$\delta$-ray contamination in the muon beam,
\item
uncertainty in the relative calibration of the sensors,
\item
alignment error of the muon monitor,
\item
effect from the tilted beamline.
\end{enumerate}
The first three sources cause a distortion in 
the observed beam profile
and lead to an uncertainty in the beam direction.
The error in the alignment 
between the target and muon monitor causes the error on the beam direction.
The beam-axis is tilted by 3.637~degrees 
downward and this results in an asymmetric profile at the muon monitor.
A correction factor was estimated using the MC simulation
for the measurement of the vertical direction
to account for this.

%

\subsubsection{Profile distortion\label{subsubsec:profile_distortion}}

The muon beam profile is reconstructed by fitting the collected charge distribution
assuming it has a form of perfect Gaussian.
However, the profile can be deviated from the ideal Gaussian form
reflecting the geometrical shape of the upstream materials.
In addition, the secondary particle, such as $\delta$-ray,
generated at the nearby materials could further distort the observed beam profile.

\begin{figure}[tb] 
   \centering
   \includegraphics[width=1.\textwidth]{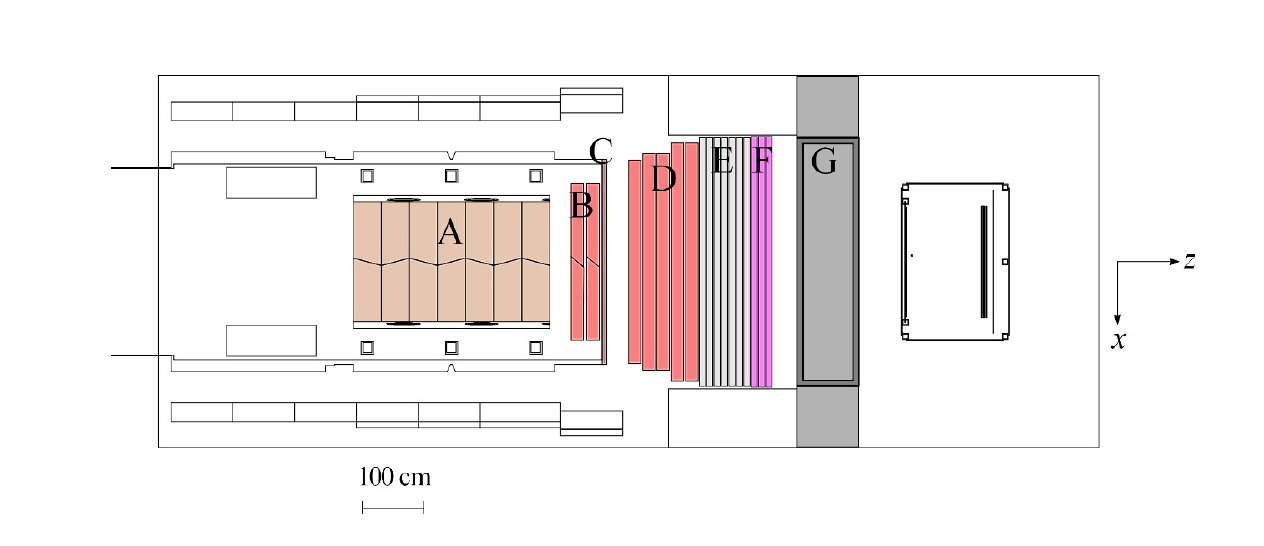}
   \caption{Top view of the beam dump. A: graphite core. B-F: Fe plates. G: concrete wall.
   The beam enters from the left side.}
   \label{fig:beamdump}
\end{figure}

The beam dump consists of multiple objects as shown in Fig.~\ref{fig:beamdump}.
The deviation of the thickness or density of objects from the design value
causes a non-uniformity and may distort the muon profile.
The effect was estimated with a MC simulation.
In the simulation,
both the density ($\rho$) and thickness ($d$) 
of the one half (positive side in the horizontal direction) of the 
each dump component
are adjusted up/down by their listed errors.
Table~\ref{tab:beamdump} 
summarizes the adjusted
components 
of the beam dump and 
the resultant shifts of the profile center.
In total, 
a value of 0.38~cm (0.032~mrad) was assigned 
to the systematic error of the profile center (beam direction).

\begin{figure}[tb] 
   \centering
   \includegraphics[width=1.\textwidth]{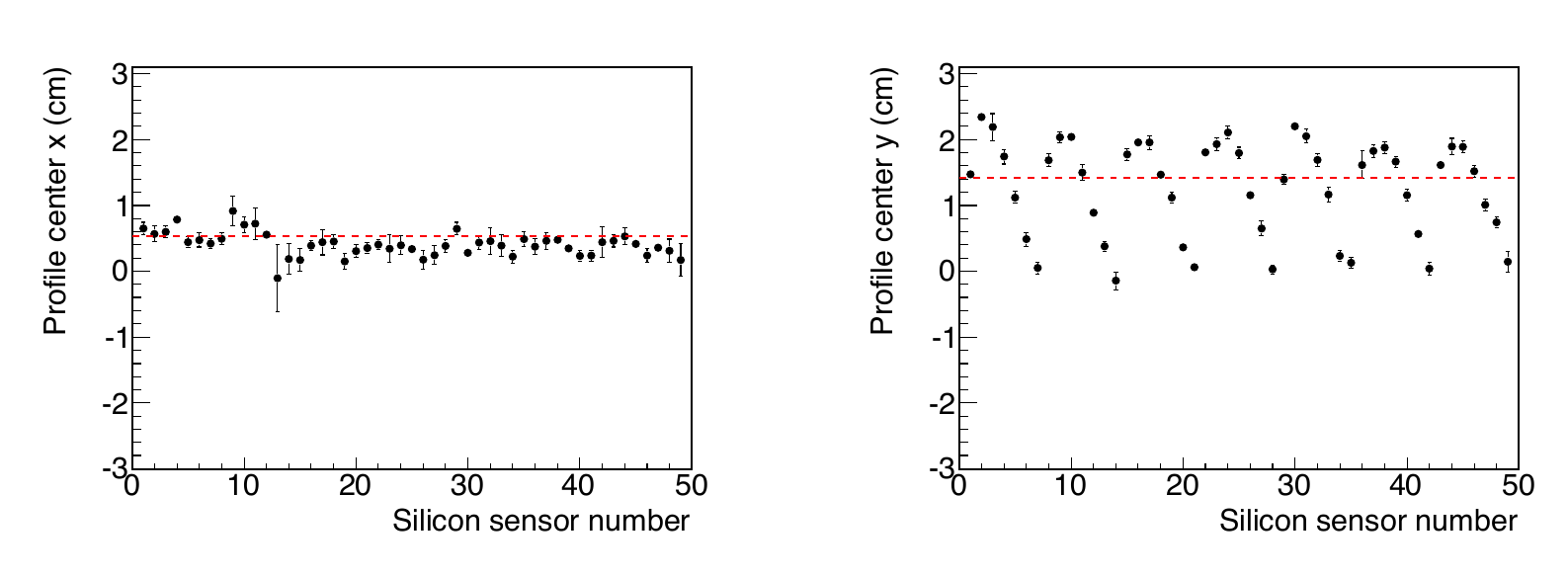} 
   \caption{Variation of the muon profile center measured by the chamber array 
   during the calibration of the silicon PIN photodiode.
   The dashed line shows the profile center for the same beam condition 
   when the silicon moving stage is lowered at the bottom (nominal position).}
   \label{fig:sicalib_iccenter}
\end{figure}

The other source of distortion of the profile 
is soft secondary particles ($\delta$-rays and $\gamma$s) from nearby materials.
We have evaluated the effect of profile distortion due to surrounding materials in two ways.
The muon monitor covers an area of $150\times150$~cm$^2$ 
transverse to the beam-axis
while the actual profile width (1$\sigma$) of the muon beam is typically 100-110~cm at the monitor
when the horns operate at 250~kA.
Namely, the muon monitor covers $\sim$50\% of the profile region.
In order to check how the actual profile deviates from the ideal Gaussian shape,
the ionization chamber arrays were moved by $\pm$25~cm to take the tail of the 
profile into account.
Then, the fit was done for the different portions of the same profile.
If the profile has a perfect Gaussian shape,
the fitted result will always be same at different positions of the array.
However, the result showed that there are differences 
in the fitted results.
The maximum differences among the fitted profile centers are
1.25~cm (0.106~mrad) for the horizontal direction 
and 1.12~cm (0.095~mrad) for the vertical direction.
During beam operation, a discrepancy of the profile center 
has been observed between the chamber and silicon arrays
(0.55~cm in the horizontal direction and 1.77~cm in the vertical direction).
This discrepancy is considered to be due to 
the difference of the nearby 
structures in-between the chamber and silicon arrays,
causing the profile to be distorted differently at the chambers and silicon arrays.
The most probable structure that would 
cause the discrepancy is the silicon moving stage just behind the silicon array
(see Fig.~\ref{fig:si_move_stand}). 
Figure~\ref{fig:sicalib_iccenter} shows 
the profile center measured by the chamber array
during the calibration of the silicon array
where the stage was moved to be positioned at each of the silicon sensors.
When the silicon moving stage cornering on the top sensors,
the profile center in the vertical direction measured by the chamber array 
shifts by $-1.4$ cm from 
the nominal case in which the stage is lowered to the bottom.
This suggests nearby materials affects the beam profile at the muon monitor.
The shift observed in two condition, 
1.25~cm for different chamber array positions
and 1.77~cm for difference in the profile center between the silicon and chamber arrays,
are taken as the systematic error.
Even though part of the shift may be caused by the dump core structure,
we conservatively add these errors since 
we cannot distinguish the effects.

As discussed in Sec.~\ref{subsec:calibration},
gain of the sensors are relatively calibrated with a precision of 0.4\% for the ionization chambers
and 0.1\% for the silicon PIN photodiodes.
The uncertainty in this calibration was propagated to the error in the beam direction.
As a result, 0.08~cm (0.007~mrad) and 0.30~cm (0.026~mrad) for the
horizontal and vertical directions respectively, were
assigned to the systematic error for the profile center (beam direction).

In conclusion, the total systematic error in the beam direction due to the
profile distortion was estimated to be 
2.20~cm in the horizontal direction and 2.22~cm in the vertical direction.
These correspond to 0.187~mrad (horizontal) and 0.188~mrad (vertical) beam direction errors.

\begin{table}[tb]
   \centering
      \caption{Density ($\rho$), thickness ($d$) and their uncertainties of the dump graphite core, Fe plates and concrete wall
      (see Fig.~\ref{fig:beamdump}). The shift of the profile center is estimated 
        for the case $\rho$ and $d$ of one half of each component are adjusted by their errors.}
   \begin{tabular}{l l l l l }
	\hline
	\hline
	& Material & $\rho$ (g/cm$^3$) & $d$ (cm) & Profile center shift (cm) \\
	\hline
	A & Graphite & $1.707\pm0.009$ & $45.001\pm0.003$ & Negligible \\
	B & Fe & $7.83\pm0.03$ & $20.00^{+0.24}_{-0.12}$  & 0.107 \\ 
	C & Fe & $7.85\pm0.02$ & $8.00^{+0.17}_{-0.09}$ &  0.054 \\
	D & Fe & $7.83\pm0.03$ & $20.00^{+0.24}_{-0.12}$ &  0.169 \\
	E & Fe & $7.8435\pm0.0083$ & $10.083\pm0.033$ & 0.083 \\
	F & Fe & $7.85\pm0.02$ & $10.00^{+0.23}_{-0.11}$ & 0.126 \\
	G & Concrete & $2.30\pm0.023$ & 100  & 0.276 \\
	\hline
	& & & Total & 0.38 \\
	\hline
	\hline
   \end{tabular}
   \label{tab:beamdump}
\end{table}

\subsubsection{Effect of the tilted beamline against the beam dump}

The beamline is tilted by 3.637~degrees vertically
while the level of the beam dump is even with the ground.
(see Fig.~\ref{fig:secondary_beamline}).
This results in asymmetric path lengths of the muons going through the beam dump
with respect to the beam-axis.
Thus, an asymmetric profile of the muon beam is observed at the muon monitor.
This causes 1.35~cm profile center shift in the vertical direction, which
was estimated using MC simulation 
where the center of the proton beam was set to the center of the target
and parallel to the beam-axis.
This shift is
 used for the correction in the beam direction measurement
and a MC statistical error of 0.22~cm (0.019~mrad) was assigned 
to the systematic error of the profile center (beam direction).


\subsubsection{Alignment error of the muon monitor}

For the systematic error in the beam direction measurement,
alignment accuracy between the muon monitor and target 
is also taken into account.
The alignment error mainly comes from the measurement error of the relative position of reference points
between the target and muon pit,
determined to be
 6.1~mm for the horizontal position and 6.3~mm for the vertical position.
In addition, 
alignment error also comes from the setting of the muon monitor (1~mm) and the target ($<1$~mm).
The total alignment error of the muon monitor relative to the target is 
therefore 6.3~mm (horizontal) and 6.5~mm (vertical).
Thus, the systematic error of the muon beam direction was 0.054~mrad in the horizontal direction and 0.055~mrad in the vertical direction.

\subsubsection{Summary of the systematic error on the beam direction measurement}

Table~\ref{tab:summary_systematics} summarizes 
the systematic error for each source.
Some of these systematic errors may come from the same origin, 
but we conservatively take quadratic sum of these as the total systematic error.
The total systematic error on the measurement
of the beam direction was estimated to be 
0.28~mrad ($=\sqrt{0.19^2 + 0.20^2}$).
Thus, the performance of the muon monitor fulfills our requirement of 0.3~mrad.

\begin{table}[tb]
   \centering
      \caption{Summary of the systematic error for the beam direction measurement.}
   \begin{tabular}{l l l l l }
   \hline
   \hline
   Error source & \multicolumn{2}{l}{Profile center} & \multicolumn{2}{l}{Beam direction} \\
		& $\Delta x$ (cm) & $\Delta y$ (cm) & $\Delta \theta_x$ (mrad) & $\Delta \theta_y$ (mrad) \\
   \hline
   Profile distortion & 2.20 & 2.22 & 0.187 & 0.188 \\
   Tilted beam & $-$ & 0.22 & $-$ & 0.019 \\
   Alignment & 0.63 & 0.65 & 0.054 & 0.055 \\
   \hline
   Total & 2.3 & 2.3 & 0.19 & 0.20 \\
   \hline
   \hline
   \end{tabular}
   \label{tab:summary_systematics}
\end{table}


\section{Measurement of the beam direction and beam tuning with the muon monitor\label{sec:operation}}
\begin{figure}[tb] 
    \centering
    \includegraphics[width=1.\textwidth]{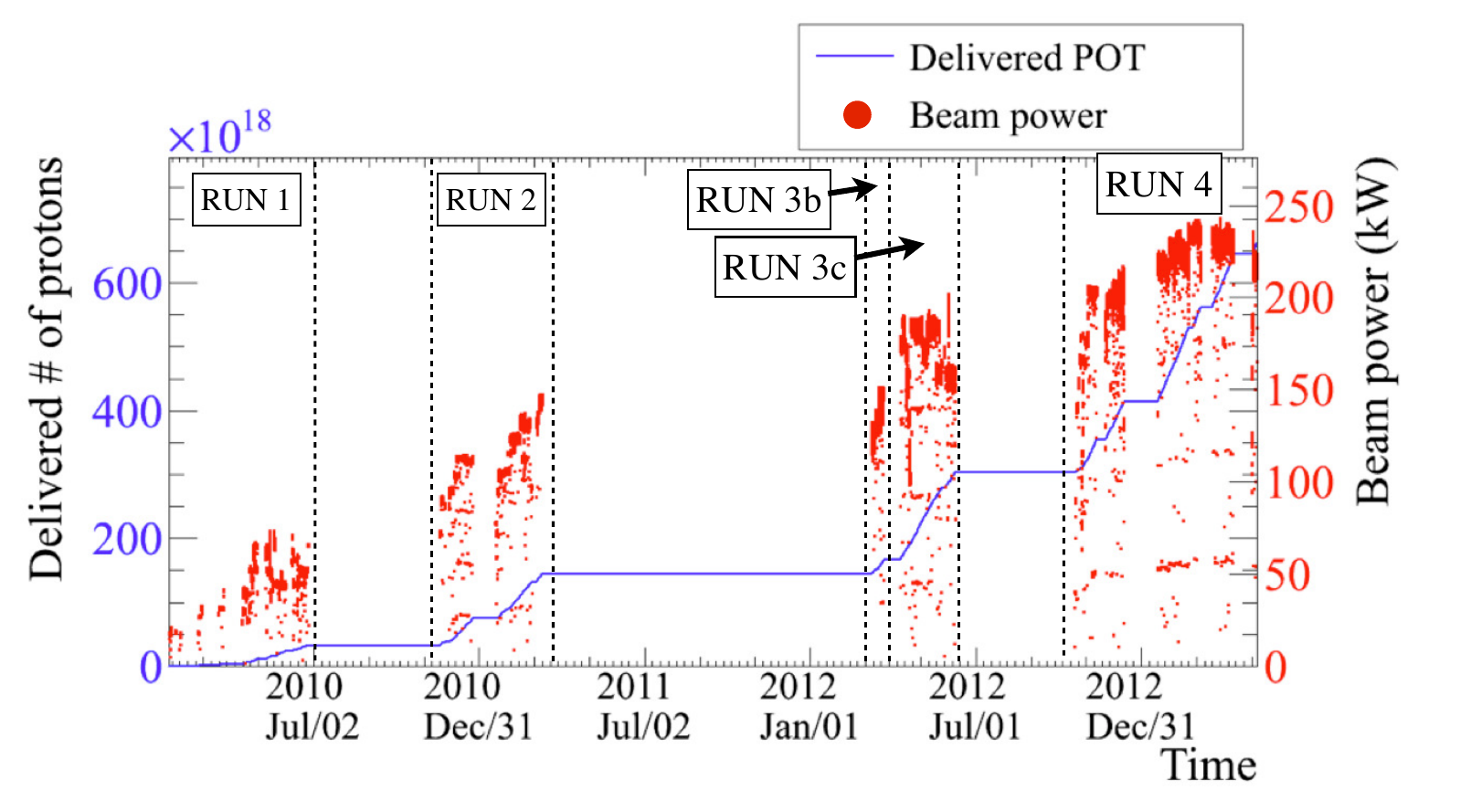} 
    \caption{History of total accumulated protons and beam power. 
    The solid line shows the accumulated p.o.t.
    The dot points show the beam power.}
    \label{fig:pot}
 \end{figure}

Table~\ref{tab:operation} summarizes 
the status of T2K beam operation 
since the start of physics data taking in January 2010.
There have been four data taking periods (RUN~1-4) until May 2013.
The repetition cycle of the proton beam
has been reduced over the course of beam operations
and 2.48~s was achieved for RUN~4.
All three magnetic horns
were operated at 250~kA
except for RUN~3b where the horns were
operated at 205~kA.
Figure~\ref{fig:pot} shows the history of the 
total accumulated p.o.t., 
as well as the beam power.
The beam power has increased gradually
and reached 220~kW ($1.40\times10^{13}$~p.p.b. with a 2.48~s repetition cycle) during RUN~4.
The muon monitor 
plays an important role 
in measuring the direction and intensity of the muon beam as described in Sec.~\ref{subsec:stability}.
At the commissioning stage of the experiment,
the horn currents were varied from 0~kA to 250~kA to 
check the dependence of the muon flux on the horn current.
We also varied the currents by $\pm$1\% from $\sim$250~kA 
in order to check if the muon monitor is sensitive to this level of variation 
in the horn currents.
In addition, the monitor was used as a tool for a survey of the components in the secondary beamline--this result was useful for understanding the current configuration of the baffle and target.
Details of the measurement are provided in Sec.~\ref{subsec:survey}.
Finally, section~\ref{subsec:horn_discrepancy} describes the 
property of the muon beam direction with 205~kA operation.

\begin{savenotes}
\begin{table}[tb]
\centering
\caption{Summary of the status of the beam operation in T2K.
The second and third column shows the repetition cycle of the proton beam
and the horn current, respectively.
The accumulated number of p.o.t. obtained for each run is shown in the last column.}
\begin{tabular}{l l l l l}
\hline
\hline
& Period & Rep. cycle (sec)  & Horn curr. (kA) &  Accum. p.o.t. \\
 \hline
 RUN~1 & Jan. 2010 $-$ Jun. 2010 &  3.52 & 250  & $3.28\times10^{19}$\\
 RUN~2 & Nov. 2010 $-$ Mar. 2011 &  3.2 \footnote{3.04~sec from March 7th 2011 to March 11th 2011}& 250 & $1.12\times10^{20}$\\
 RUN~3b & Mar. 2012 & 2.92 & 205 & $2.15\times10^{19}$\\
 RUN~3c & Apr. 2012 $-$ Jun. 2012 & 2.56 &  250 & $1.37\times10^{20}$\\
 RUN~4   & Oct. 2012 $-$ May 2013 & 2.48 & 250 & $3.60\times10^{20}$\\
 \hline
 & & & Total & $6.63\times10^{20}$\\
 \hline
 \hline
 \end{tabular}
 \label{tab:operation}
 \end{table}
 \end{savenotes}

 \subsection{Proton beam tuning with the muon monitor\label{subsec:beam_tuning}}
\begin{figure}[tb] 
    \centering
    \includegraphics[width=1.\textwidth]{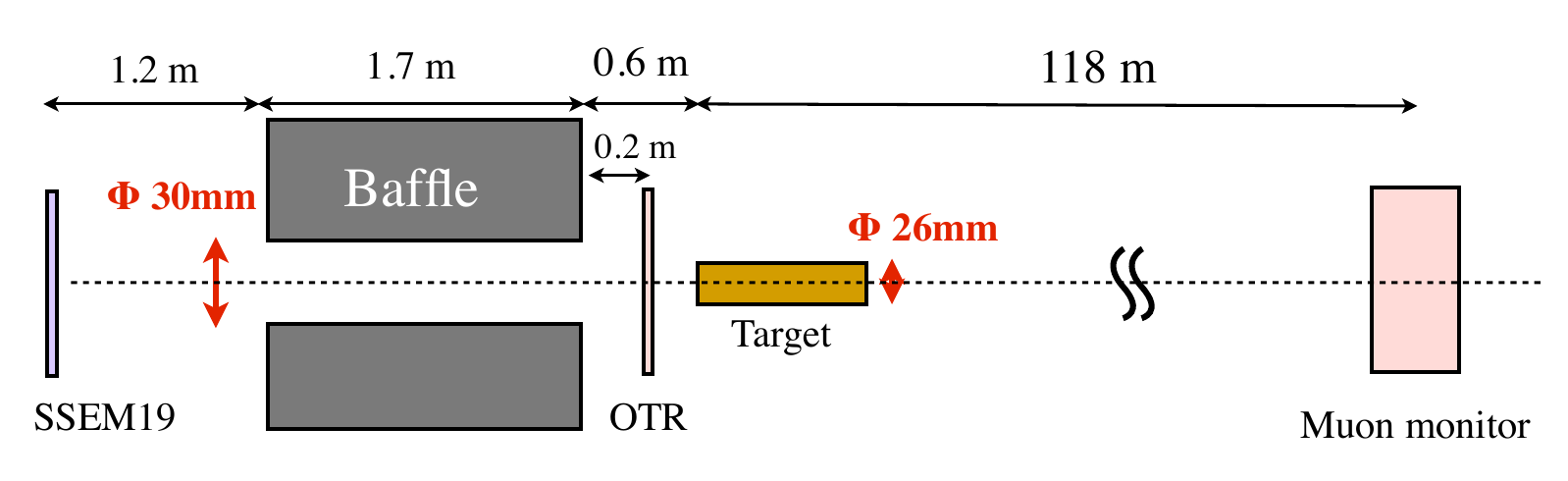} 
   \caption{Plan view configuration of the components in the secondary beamline with SSEM19 and the muon monitor.}
    \label{fig:beam_alignment}
 \end{figure}

\begin{figure}[tb] 
    \centering
    \includegraphics[width=1.\textwidth]{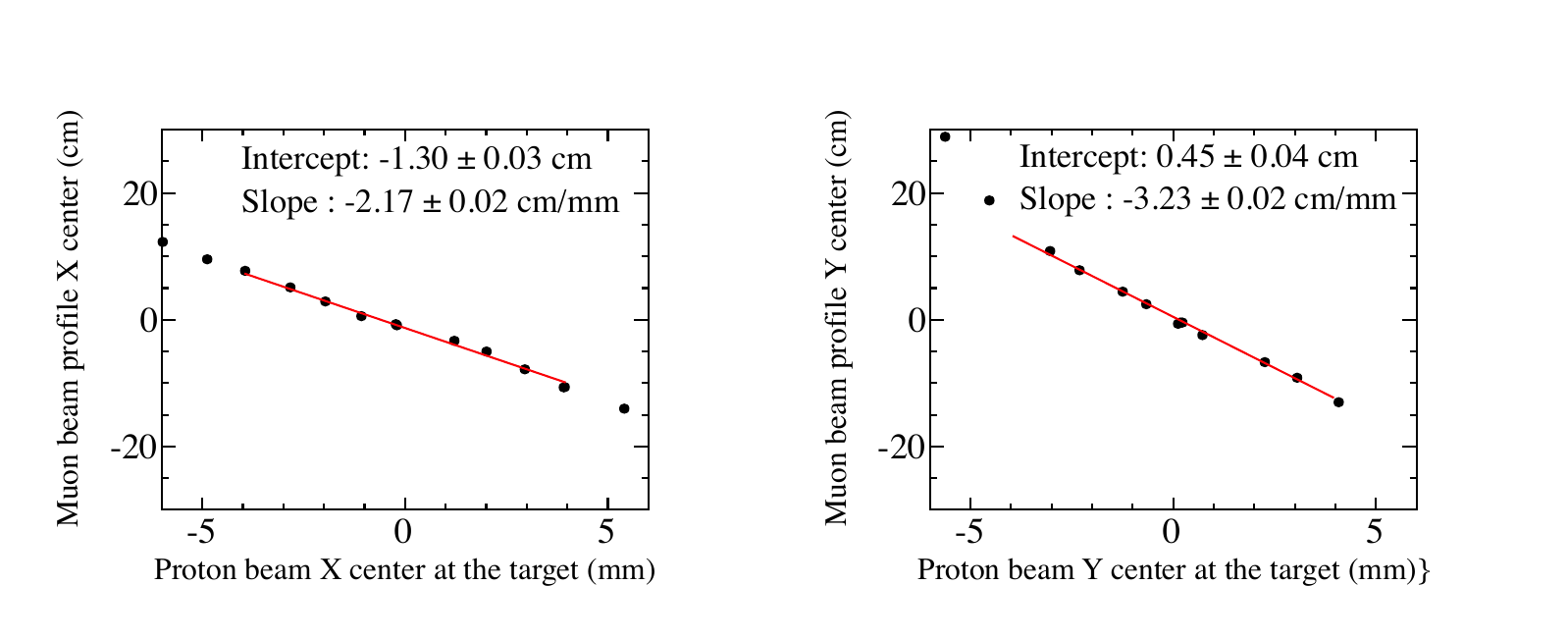} 
    \caption{ Correlation between the profile center at the muon monitor and the proton beam
    position at the target
    in the horizontal (left) and vertical (right) direction. 
    The position of the proton beam is
     extrapolated at the target using the measurement of the SSEMs and OTR.
     Fitted lines (red lines) and the results are shown in the figures. 
     The errors on the fitted parameters
      are only statistical ones. }
    \label{fig:pbeam_scan}
 \end{figure}

 Figure~\ref{fig:beam_alignment} shows a schematic view of the configuration of 
the components in the secondary beamline, 
the proton-beam monitor, 
as well as the muon monitor.
Before hitting the target, the proton beam 
passes
 thorough proton-beam monitors
placed in the primary beamline 
just upstream of the secondary beamline.
Segmented secondary emission profile monitors (SSEM)
are used for monitoring the profile center and width of the proton beam.
The baffle is placed downstream of SSEM19,
which is the most downstream SSEM
 and this plays the role
of a collimator 
with an opening of 30~mm.
An optical transition radiation (OTR) monitor~\cite{otr} is placed just in front of the target
and is used for the measurement of the position of the proton beam. 
Using the SSEMs and OTR measurements, the
beam position upstream 
of the baffle (target)
is reconstructed with an accuracy better than 0.7~mm (0.6~mm).
Data at various beam position were taken to 
measure the correlation between 
the proton beam position at the target and the profile center 
of the muon beam at the muon monitor.
 As shown in Fig.~\ref{fig:pbeam_scan},
 the profile center measured by the muon monitor 
 is very sensitive to the position of the proton beam
 at the target.
The angle and position of the proton beam are tuned 
very precisely using the correlation
such that the profile center of the muon beam is centered at the muon monitor.
 
\subsection{Stability of the beam direction and intensity measured by the muon monitor\label{subsec:stability}}

\begin{figure}[tb] 
    \centering
    \includegraphics[width=1\textwidth]{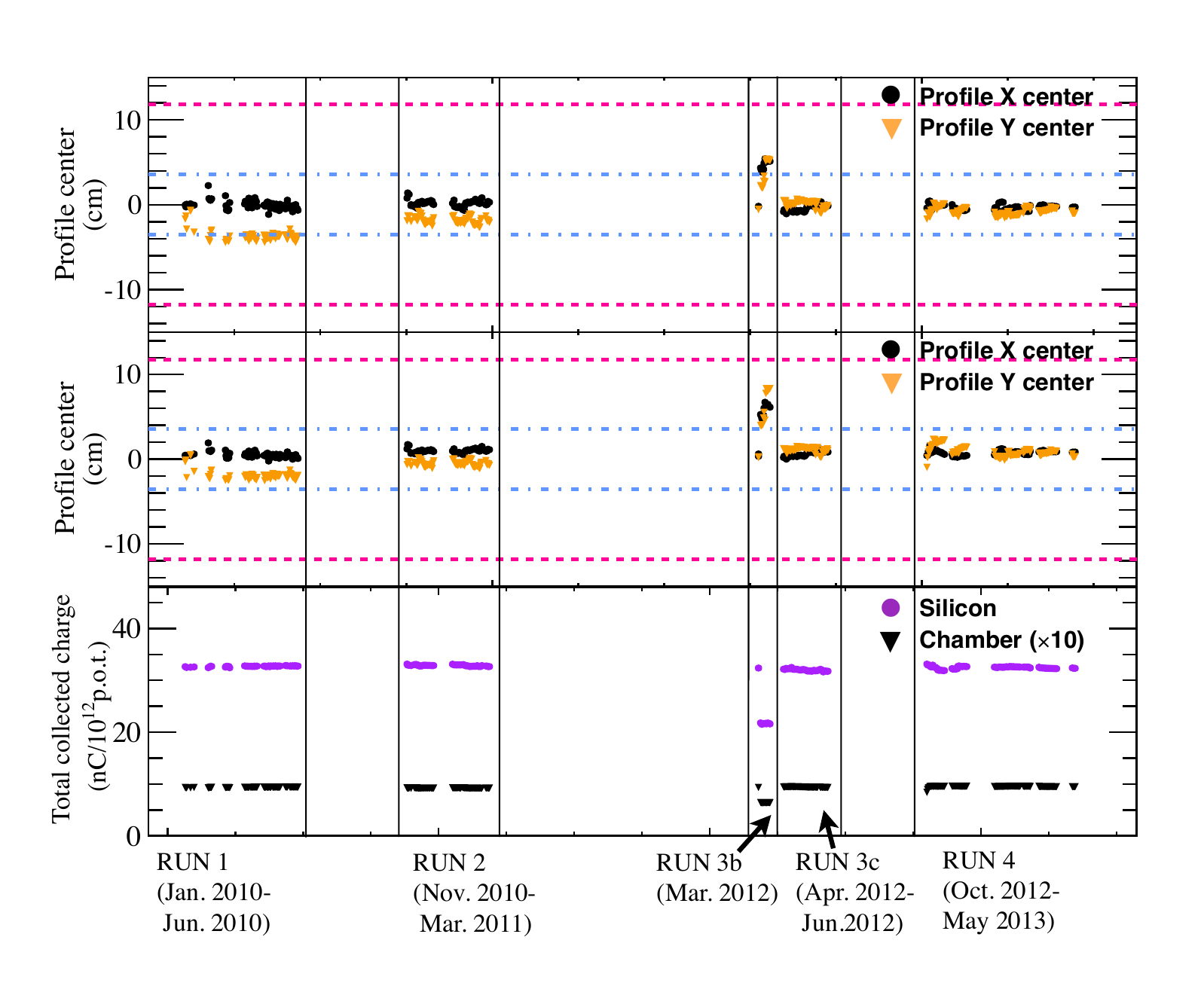} 
    \caption{Daily stability of the muon beam measured by the muon monitor.
The profile center measured by the silicon and chamber array is shown in the top
and middle respectively.
The profile center values corresponding to a beam direction of 1.0~mrad and 0.3~mrad
 are also displayed as pink and blue dashed lines, respectively.
The total collected charges measured by the silicon and chamber
(scaled up by 10 times) are shown in the bottom of the figure.}
    \label{fig:stability}
 \end{figure}

\begin{figure}[tp] 
    \centering
    \includegraphics[width=1.0\textwidth]{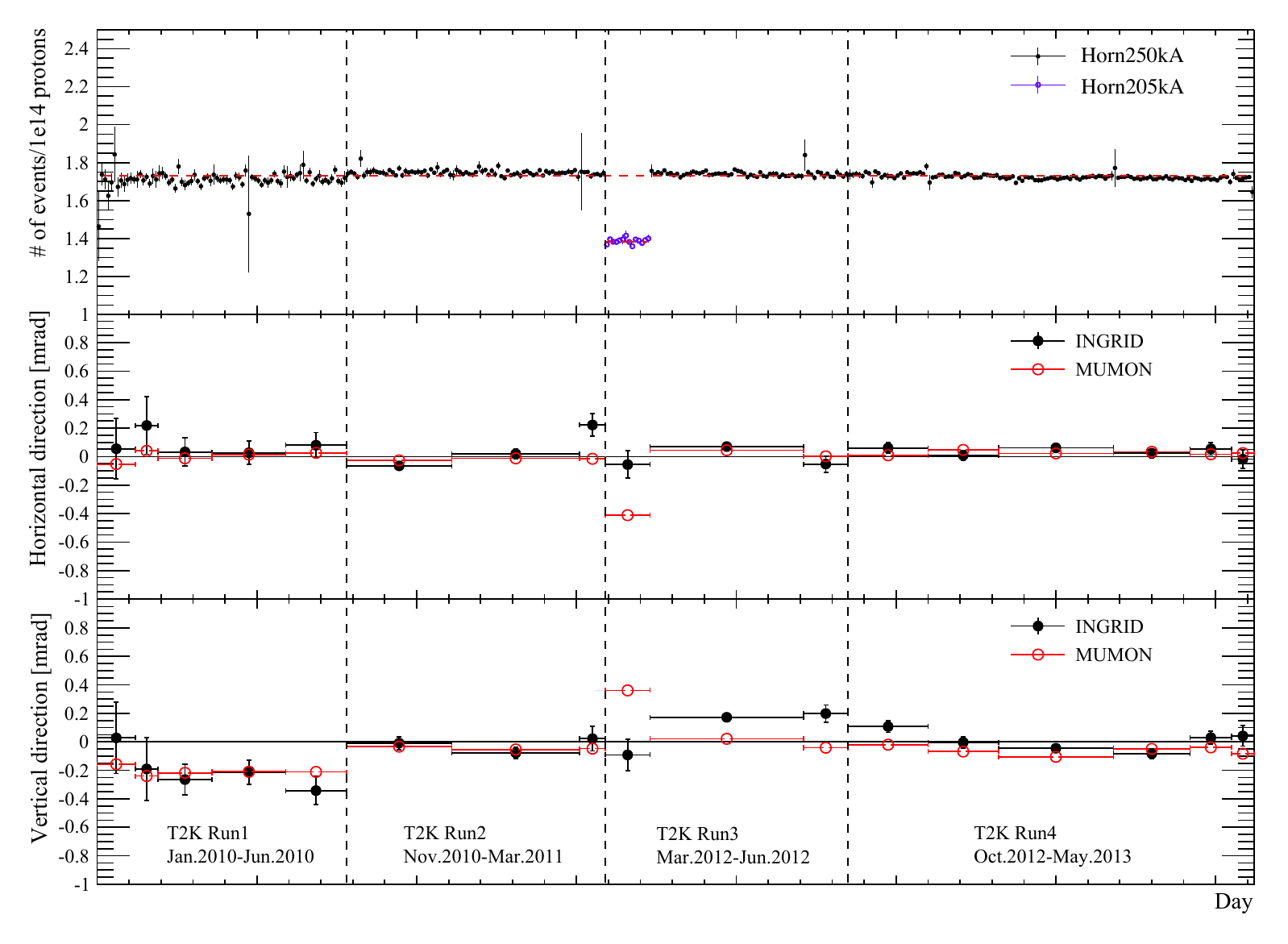} 
    \caption{Top: Neutrino event rate per $10^{14}$~p.o.t.
    measured by INGRID (points) with the mean value (dashed line).
    Middle and Bottom: Beam direction measured by the muon monitor (red open circle)
    and INGRID (black circle) in the horizontal and vertical direction respectively.
    The error bar represents the statistical error.
    In this figure, a sign is reversed for the horizontal direction measured by the muon monitor so that the x-coordinate for the muon monitor matches that for INGRID.  }
    \label{fig:stability2}
 \end{figure}

As described in Sec.~\ref{subsec:beam_tuning}, 
the proton beam is tuned using information from the muon monitor
and always controlled 
such that 
the muon beam and hence the neutrino beam are on the beam-axis.
Stability of response of silicon sensors is confirmed as described below.
As a radiation dose at the sensor on beam-axis 
is expected to be twice as high as that at the edge sensor,
degradation due to the radiation damage 
is expected to be different between these two sensors.
The effect of different radiation dose can be 
estimated using calibration data taken in different periods,
where signals from the sensors are measured at various positions (see Sec.~\ref{subsec:calibration}).
The difference in the signal size between the edge and center sensors was checked for different calibration data sets. 
We then checked if there is any significant decrease in the signal size at the center sensor where the radiation dose is highest.
As a result, we confirmed that there was no significant decrease 
in the signal size after accumulating $\sim1.0\times10^{20}$~p.o.t.
From this result, we ensure that 
the response of the sensor is stable for operation accumulating $\sim1.0\times10^{20}~$p.o.t., which is a typical value 
of p.o.t. obtained in each run period (Run~1-3).
For Run~4 operation where more than $1.0\times10^{20}$~p.o.t. was accumulated, we rely on the 
result from the beam test and assumption discussed in Sec.~\ref{subsec:Si}.
Figure~\ref{fig:stability} 
shows the daily stability of the muon beam as measured by the muon monitor.
The profile center measured by the silicon and chamber array is shown in the top
and middle 
sections respectively.
As shown in the history of the beam direction,
most of the events 
lie within 0.3~mrad except for RUN~1 and RUN~3b.
After the RUN~1 operation, 
the center position of the muon monitor was 
found to be mistakenly 
mis-aligned by -2.5~cm in the vertical direction.
This mis-alignment was taken into account for the beam tuning 
from the RUN~2 operation onwards.
The magnetic horns were operated at 205~kA during the RUN~3b operation.
Unlike in
the case of 250~kA operation,
the profile center of the muon beam was 
shifted during
205~kA operation
even though the proton beam position
was tuned to the center at the target using 
the correlation shown in Fig.~\ref{fig:pbeam_scan}.
The direction and intensity of the neutrino beam
have also been measured by INGRID 
and 
this result is shown in Fig.~\ref{fig:stability2}, 
with the neutrino event rate having been stable over the majority of the run period.
The beam direction measured by INGRID
shows a different tendency from that of
the muon monitor
during the RUN~3b operation.
However, all of the spills were within 1.0~mrad for 
both the muon and neutrino beams.
In addition, most of the spills were controlled well within 
our 0.3~mrad
requirement.
Table~\ref{tab:stability} summarizes the average and RMS of the
profile center and total collected charge.
Although the profile center deviated largely during RUN~3b,
we achieved good stability in the beam direction over the entire period.
The total collected charge was also kept stable and the RMS was less than 1.0\%.

\subsection{Resolutions for the direction and intensity measurements of the muon beam\label{subsec:resolution}}
\begin{figure}[tb] 
   \centering
   \includegraphics[width=1.\textwidth]{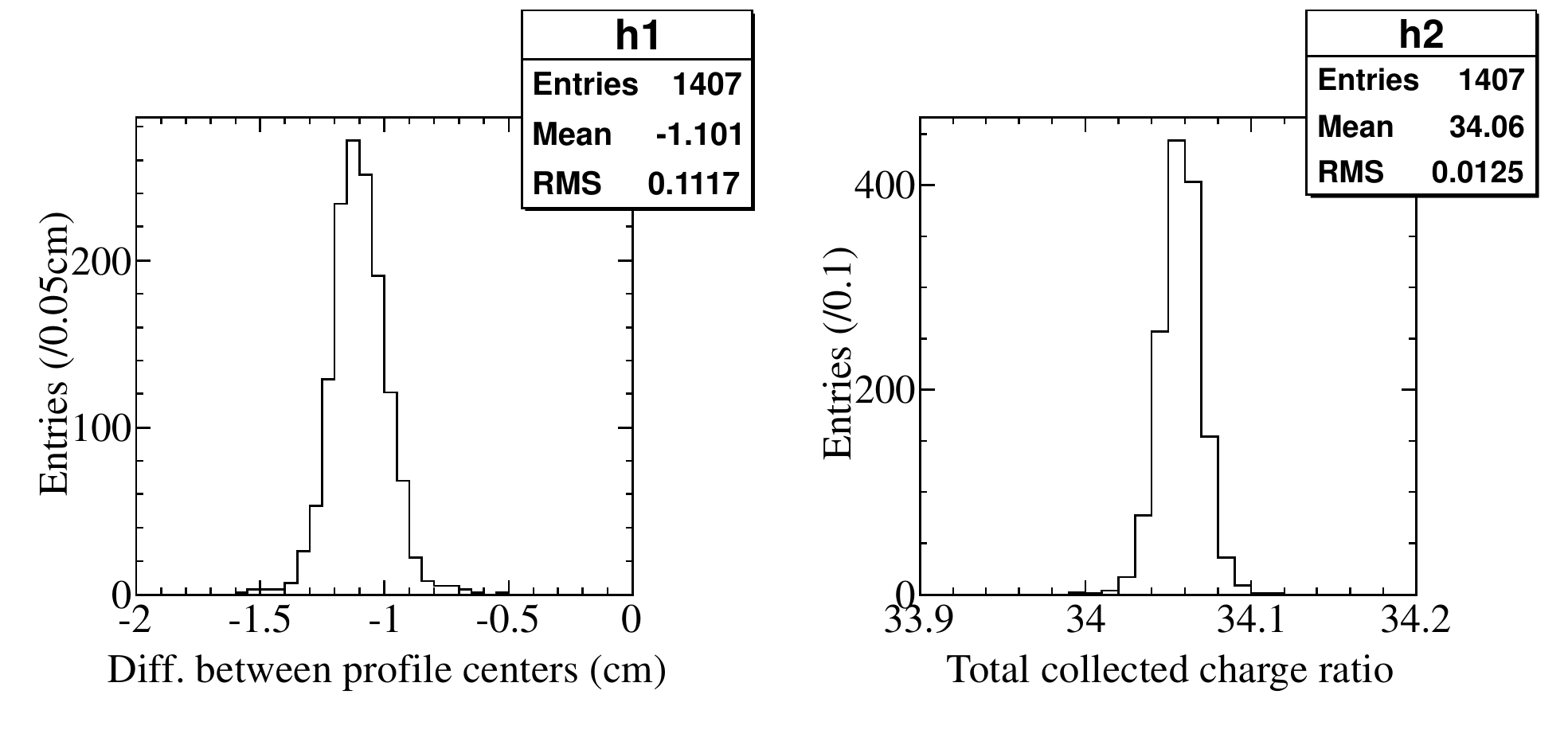} 
   \caption{
Obtained resolution of the variation in the direction (left) and intensity (right) measurement by the muon monitor. 
 For the beam direction, the difference in the measured profile center between the silicon and chamber arrays was measured.
 For the beam intensity, we took a ratio of the total collected charge measured by the silicon array to that measured by the chamber array.
 These results were obtained using beam data with 1~hour operation and $1.3\times10^{13}$~p.p.b. of the proton beam intensity. }
   \label{fig:resolution}
\end{figure}

The direction and intensity of the muon beam can 
vary spill-by-spill due to the fluctuations in the proton beam direction and in the horn current. 
The resolution 
of the variation in the direction and intensity measurement by the muon monitor
was estimated using two independent 
detectors, i.e. the silicon and chamber arrays 
in order
to reduce the effects from intrinsic beam fluctuations.
For the beam direction, we took the difference in the measured profile center
between the silicon and chamber arrays
(see left in Fig.~\ref{fig:resolution}).
For the beam intensity, we took a ratio of the total collected charge measured by the silicon array
to that measured by the chamber array
(see the right in Fig.~\ref{fig:resolution}).
Both the resolutions for the direction and intensity measurement
are actually 
convoluted signals
from the chamber and silicon sensors.
As the size of a signal from the silicon array is 30~times larger than
that from the ionization chambers, 
the resolution at lower intensity is limited by
the resolution of the ionization chambers.
The resolutions become better as the proton beam intensity increase.
As a result,
we achieve good resolutions of $<3.0$~mm for the direction
and $<0.1$\% for the intensity measurement 
when the beam intensity is above $\sim0.5\times10^{13}$~p.p.b.

\begin{table}[tb]
\centering
\caption{Average of the profile center and total collected charge measured by the muon monitor 
for each T2K run period. The numbers in parentheses denote the RMS.}
\scalebox{0.85}
{
\begin{tabular}{l | l l l | l l l}
\hline
\hline
 & \multicolumn{3}{c |}{Silicon array} & \multicolumn{3}{c}{Ionization chamber array} \\
 & \multicolumn{2}{l}{Profile center}  & Total collected charge & \multicolumn{2}{l}{Profile center} & total collected charge \\
Period & X (cm) & Y (cm) & (nC/10$^{12}$~p.o.t.)  & X (cm) & Y (cm)  & (nC/10$^{12}$~p.o.t.) \\
\hline
RUN~1 & -0.1 (0.62) & -3.8 (0.53) & 32.7 (0.7\%)  & 0.4 (0.47) & -2.0 (0.47)  & 0.939 (0.7\%) \\
RUN~2 & 0.2 (0.42) & -1.9 (0.48) & 32.8 (0.8\%)  & 1.0 (0.45) &  -0.5 (0.46)  & 0.922 (0.7\%) \\
RUN~3b & 4.8 (0.60) & 4.2 (1.52) & 21.7 (0.7\%)  & 5.9 (1.11) &  6.7 (2.18)  & 0.640 (0.7\%) \\
RUN~3c & -0.4 (0.38) & 0.1 (0.41) & 32.0 (0.7\%)  & 0.6 (0.46) &  1.1 (0.47)  & 0.942 (0.6\%) \\
RUN~4  & -0.3 (0.33) & -0.8 (0.47) & 32.4 (0.8\%)  & 0.8 (0.34) &  0.9 (0.66)  & 0.954 (1.0\%) \\
\hline
\hline
\end{tabular}
}
\label{tab:stability}
\end{table}

\subsection{Dependence of the 
muon yield
on the horn current\label{subsec:hcur_dependence}}

\begin{figure}[tb] 
   \centering
   \includegraphics[width=0.8\textwidth]{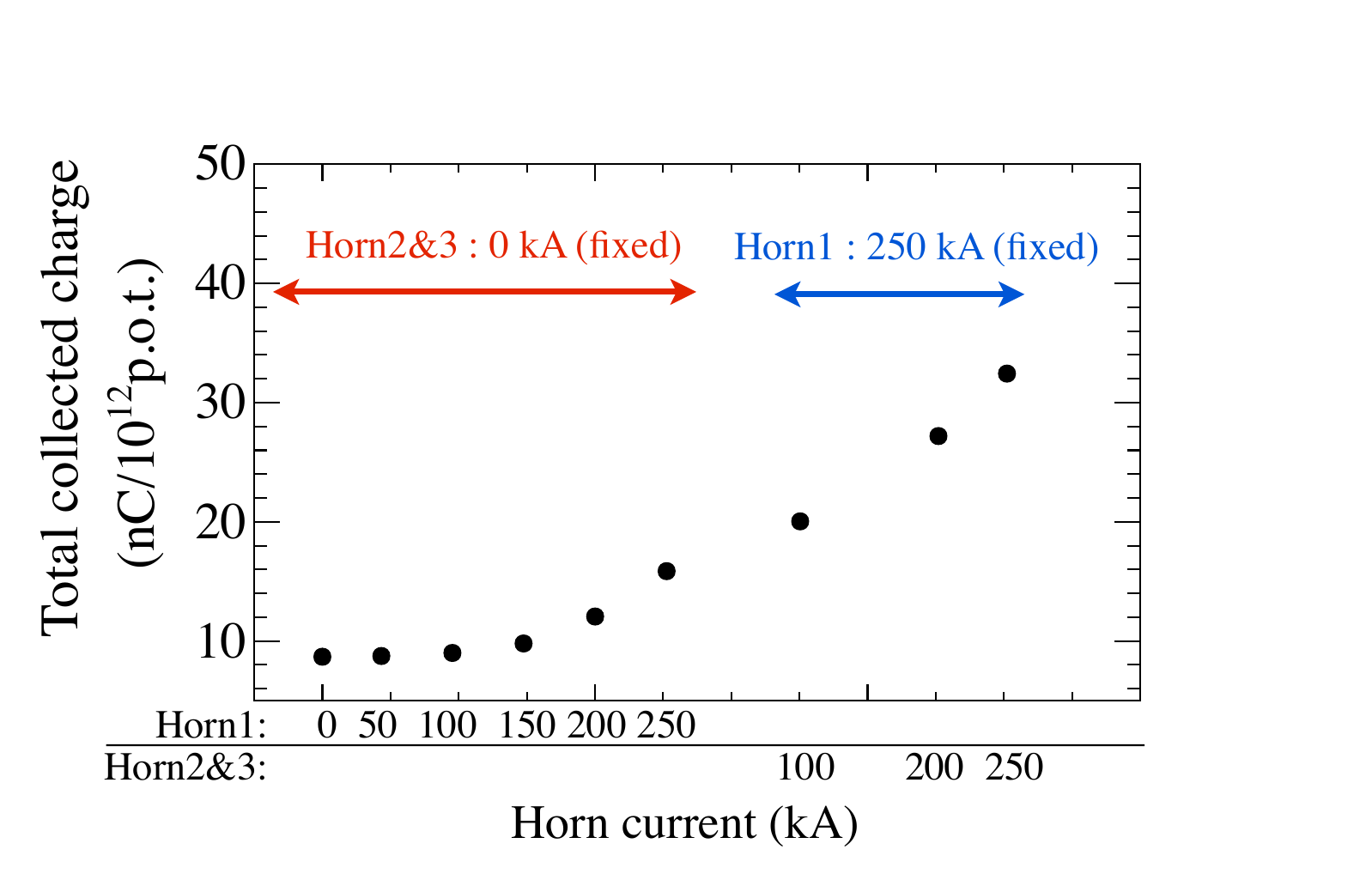} 
   \caption{Dependence of the total collected charge for different 
   combination of
   horn currents.}
   \label{fig:hcur_dependence}
\end{figure}

\begin{figure}[tb] 
   \centering
      \includegraphics[width=0.48\textwidth]{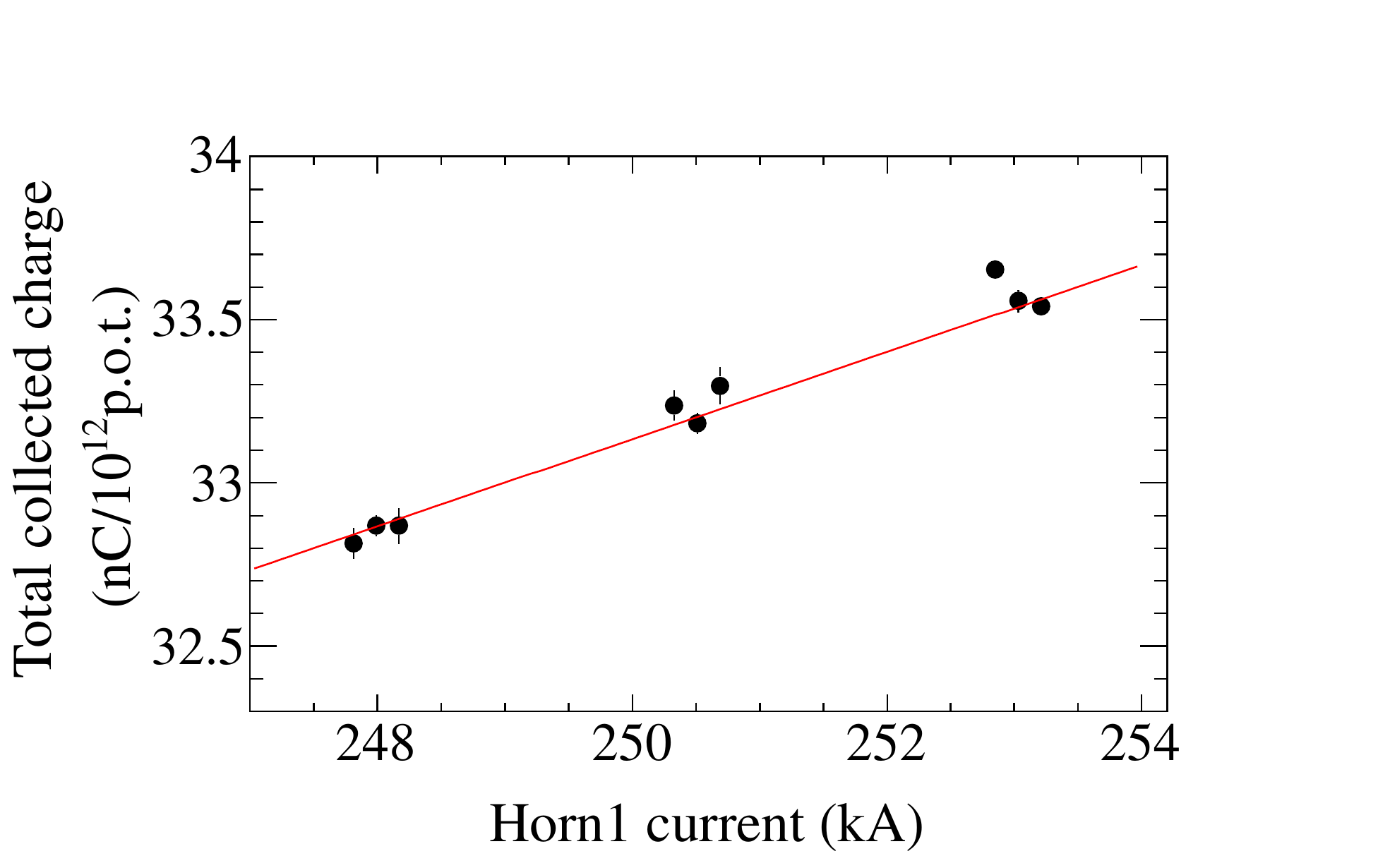} 
   \includegraphics[width=0.48\textwidth]{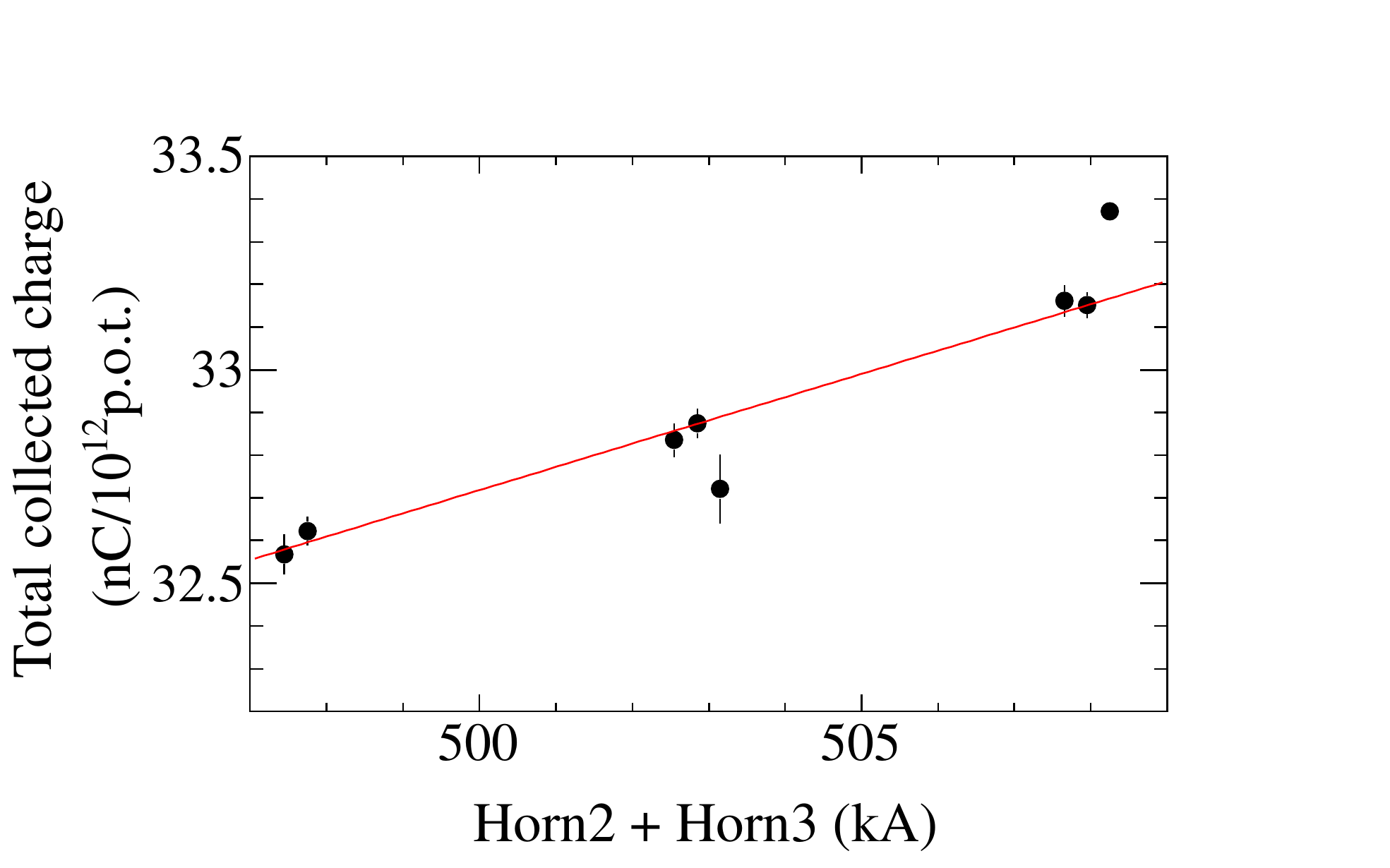} 
   \caption{Dependence of the total collected charge on the horn current variation.
  Left: the 
  Horn1
  current was changed by 
  $\pm$1\% (2.5~kA)
from the nominal value (250~kA)
while fixing the 
Horn2 and Horn3 currents at 252~kA.
Right: the 
Horn2 and Horn3 currents were 
simultaneously
changed by $\pm$1\% (2.5~kA) from the nominal values
while fixing the 
Horn1 current at 248~kA.}
   \label{fig:hcur_dependence2}
\end{figure}

Increasing the horn currents results in 
focusing more charged pions
and producing more intense muon and neutrino beams.
The focusing of pions can be confirmed by
the charge of muon flux, 
which is measured as the collected charge in the muon monitor.
During beam operation, 
we tested how the collected charge changes
by varying
 the horn currents from 0~kA to 250~kA.
Figure~\ref{fig:hcur_dependence} shows the total collected charge measured by 
the silicon array for various horn currents.
When all of the horns 
are operated at 250~kA,
the collected charge 
are increased by a factor of 4
compared with the case of 
0~kA horn current setting.
We also varied the horn current within $\pm$1\% (2.5~kA)
and checked 
the effect on the collected charge.
This result is shown in Fig.~\ref{fig:hcur_dependence2}.
When the 
Horn1 current was varied by $\pm$1~kA 
from 250~kA while fixing the 
Horn2 and Horn3 currents at 252~kA,
the collected charge measured by the silicon array
varied by 0.40\% (left in Fig.~\ref{fig:hcur_dependence2}).
Subsequently, 
Horn2 and Horn3 currents were 
simultaneously varied by $\pm$1\% from
$\sim$250~kA
while fixing the 
Horn1 current at 248~kA.
This resulted in a 0.33\%/kA change in the collected charge
(right in Fig.~\ref{fig:hcur_dependence2}).
As described in Sec.~\ref{subsec:resolution},
the muon monitor has a resolution of 0.1\% 
in the beam intensity measurement.
Thus, the monitor is sensitive to variations of $\sim$0.3~kA in 
either Horn1, or Horn2 and Horn3 combined.
These results show
that the muon monitor 
is also useful for monitoring the horn currents.

\subsection{Survey of the secondary beamline.\label{subsec:survey}}

\begin{figure}[tb] 
   \centering
   \includegraphics[width=1.\textwidth]{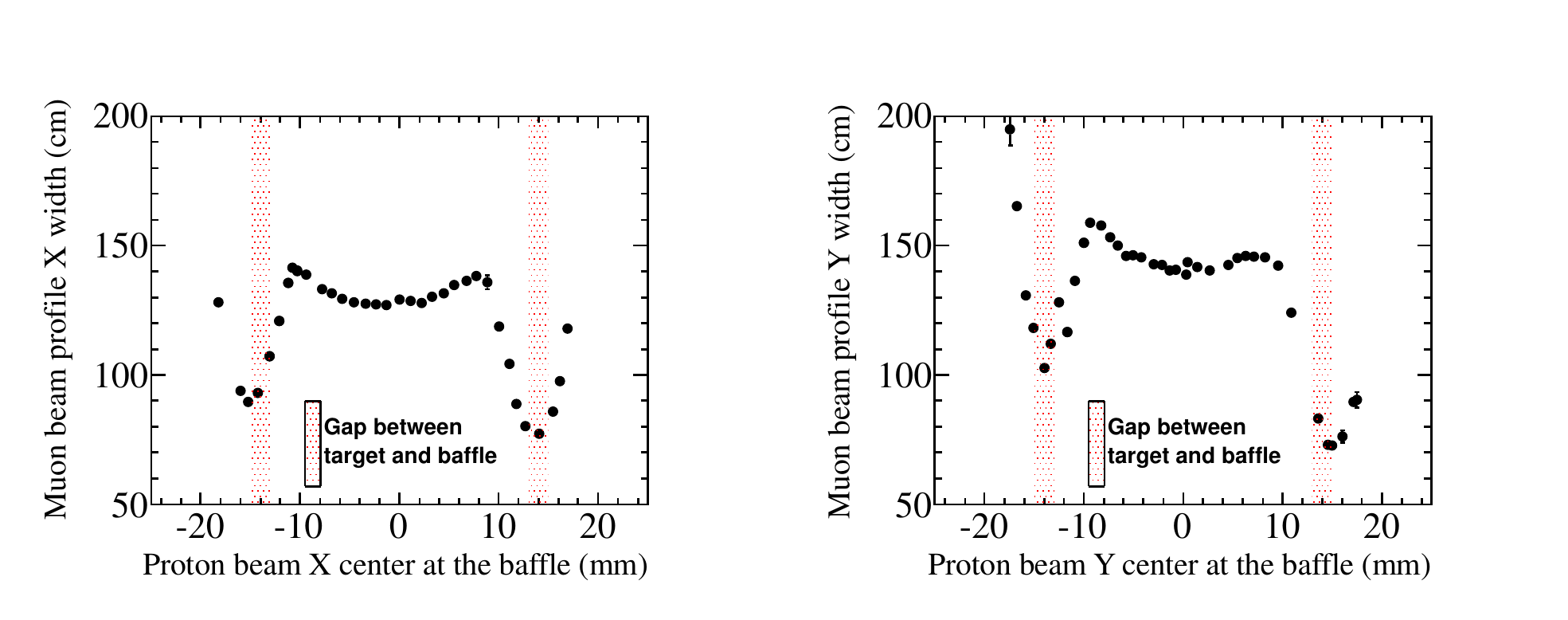} 
   \caption{Profile width of the muon beam at the silicon array
obtained by scanning the proton beam at the baffle in horizontal (left) and vertical (right).
The 
expected position of 2~mm gap
between the baffle and target is 
expressed as the red shaded region ($-15\sim-13$~mm and $13\sim15$~mm). 
All of the horn currents are set to 0~kA during 
the data taking.
}
   \label{fig:width_scan}
\end{figure}

The configuration of the components in the beamline
might be changing due to the 
sinking of the ground.
In addition, the Grate East Japan Earthquake in 2011
resulted in movement of many of the components~\cite{flux}.
The muon monitor has also played an important role 
in confirming the 
alignment of the secondary beamline.
Ideally the relative center positions should be consistent
between the baffle and target.
If there is a difference in the relative center positions
between these two components, the proton beam will hit the baffle (collimator) and will not produce secondary particles in the target 
effectively. In addition, the miss-steered beam, which is not
collimated properly by the baffle,
would result in hitting downstream components.
However, it is impossible to survey the instruments with a visual inspection during beam operation
because 
they are inside
the helium gas volume enclosed by the helium vessel.
We therefore conducted the survey using the proton beam 
during operation 
just after the recovery work for the earthquake.
The proton beam size was set 2.3-2.8~mm during
the survey run while the nominal size is $\sim$4~mm.
As shown in Fig.~\ref{fig:beam_alignment}, 
the baffle has a beam hole of 30~mm,
while the target has a diameter of 26~mm.
Namely, there is a radial
gap of 2~mm between the baffle and target.
If the alignments of these two instruments are perfect,
the proton beam 
interacts less with the target when passing
through the gap.
Then the contribution of the muons from interactions at the dump increases.
This results in a narrower muon beam at the muon monitor. 
Figure~\ref{fig:width_scan} shows 
the profile width of the muon beam at the silicon array,
obtained by scanning the proton beam position
at the baffle in horizontal (left) and vertical (right) axes.
The 
expected position of the 2~mm gap
between the baffle and target is 
expressed as the red shaded region ($-15\sim-13$~mm and $13\sim15$~mm)
in the figure.
As shown in 
Fig.~\ref{fig:width_scan},
the profile widths 
have minimums around the 
gap in both horizontal and vertical axes.
Fitting to 
these dips
with a quadratic function was 
performed
to extract the actual gap position.
The fitted dips are situated within the expected position of the 2~mm gap
between the baffle and target.


\begin{table}[tb]
\centering
\caption{Fitted gap position between the baffle and target. The error is statistical.}
\begin{tabular}{l l l l l}
\hline
\hline
 & \multicolumn{2}{l}{Scan in horizontal} & \multicolumn{2}{l}{Scan in vertical} \\
 \cline{2-5}
 & $x<0$ & $x>0$ & $y<0$ & $y>0$ \\
 \hline
 Fit result (mm) & $-15.0\pm0.04$ & $13.7\pm0.04$ & $-14.1\pm0.03$ & $14.9 \pm 0.11$ \\
 Fit range (mm) & $-17.0\sim-13.0$ & $11.5\sim15.5$ & $-16.0\sim-13.0$ & $13.0\sim16.5$ \\
 \hline
 \hline
 \end{tabular}
 \label{tab:width_scan}
 \end{table}


\subsection{Property of the muon beam direction with 205~kA operation\label{subsec:horn_discrepancy}}

\begin{figure}[tb] 
   \centering
   \includegraphics[width=1.\textwidth]{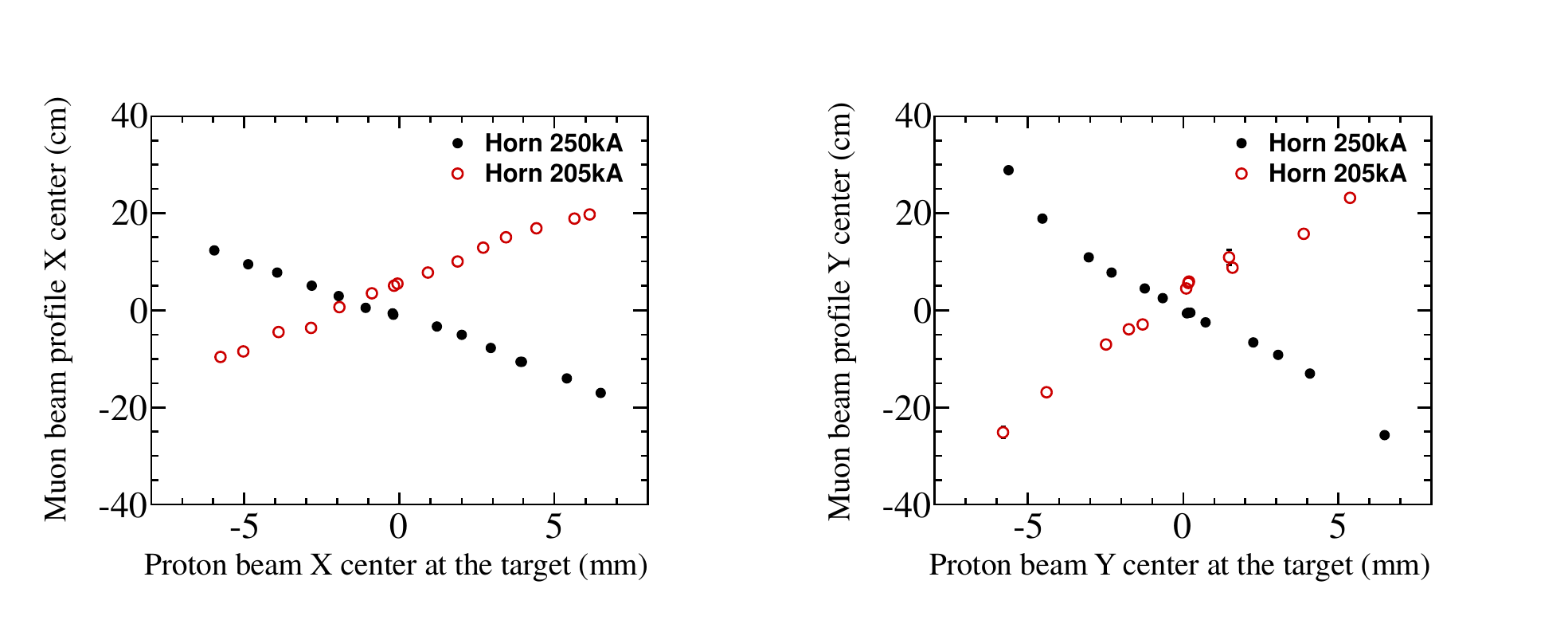} 
   \caption{Correlation between the profile center at the silicon array 
and the proton beam position at the target for the 250~kA (black) and 205~kA (red) operation.}
   \label{fig:horn_discrepancy}
\end{figure}

\begin{figure}[tb] 
   \centering
   \includegraphics[width=0.45\textwidth]{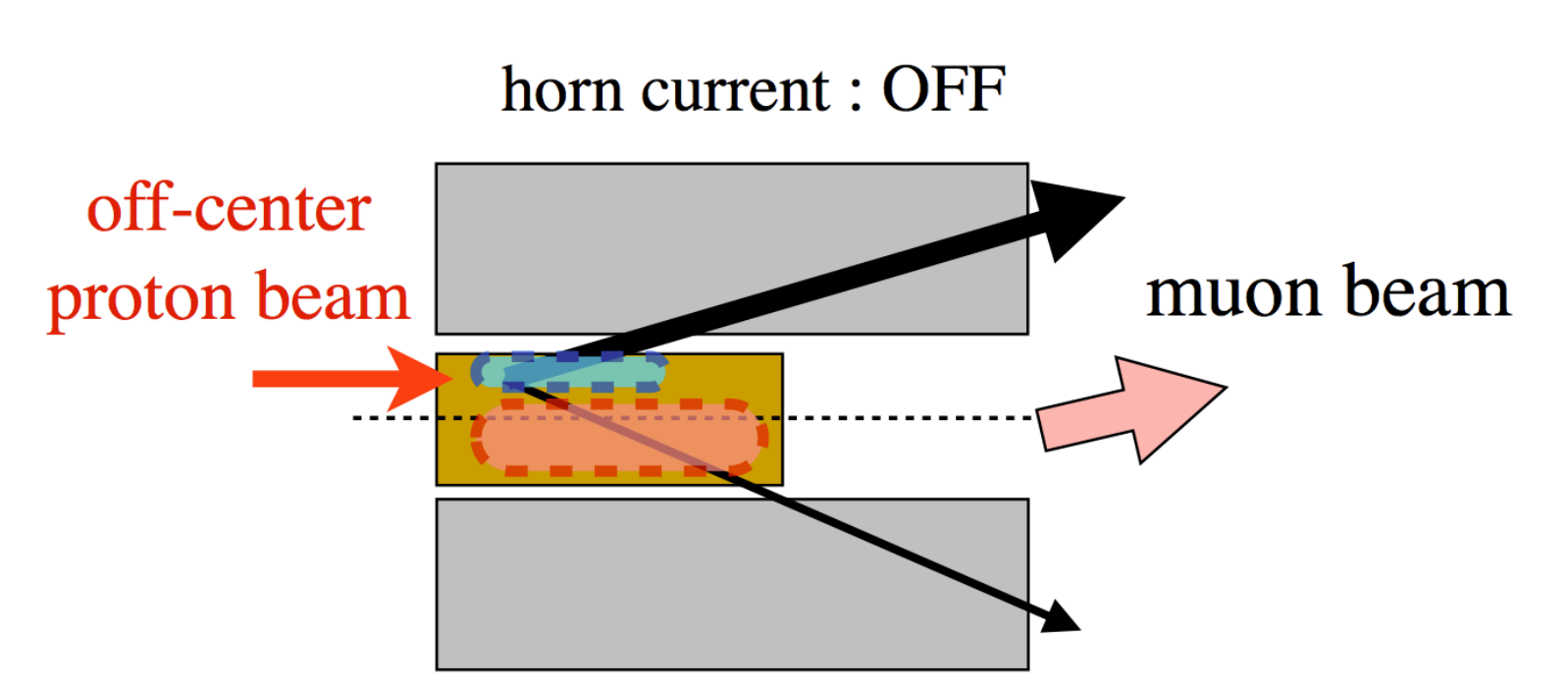}   \includegraphics[width=0.45\textwidth]{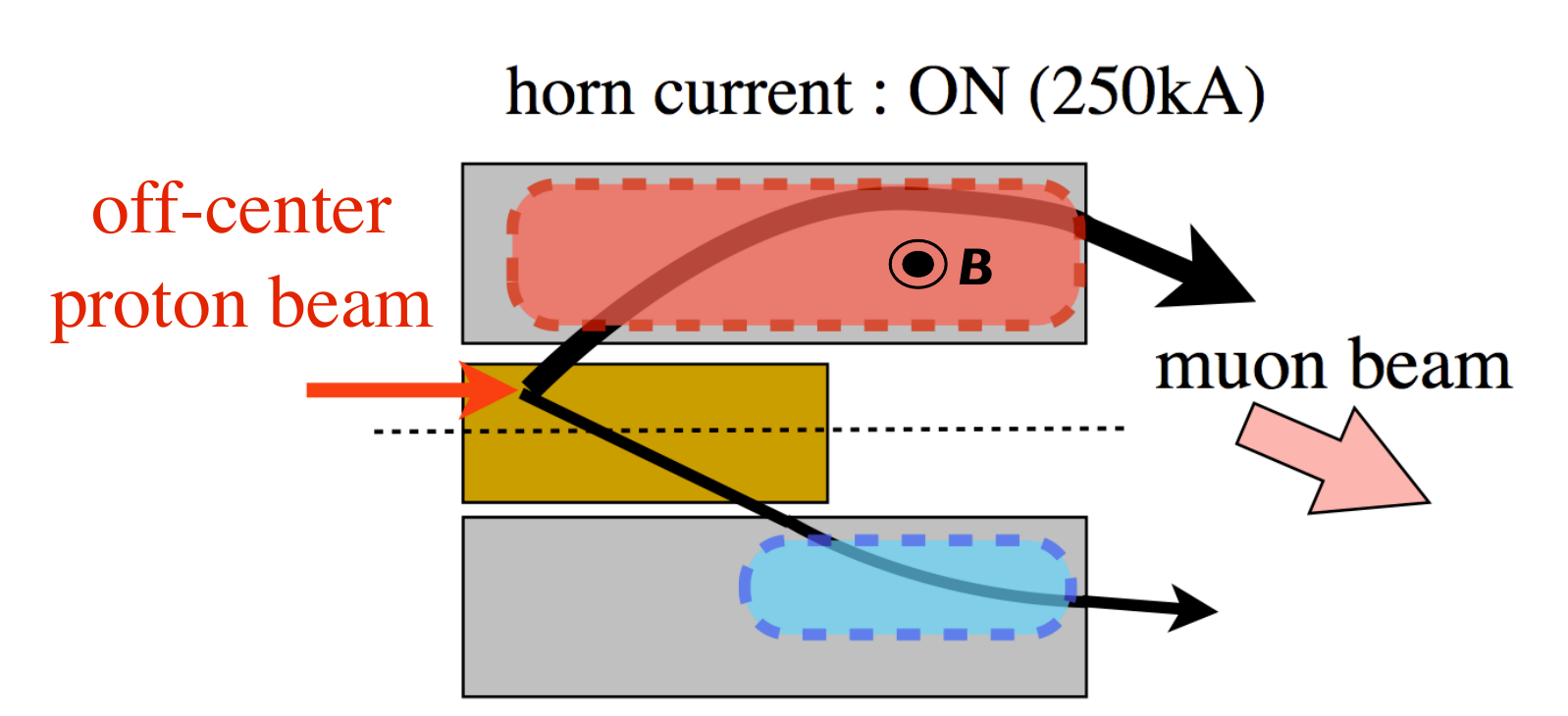} 
   \caption{Explanation of changes in the direction of the muon beam when the horn current is turned off (left) and on at 250~kA (right).}
   \label{fig:horn_focus}
\end{figure}

In order to understand property of the muon beam direction
during the 205~kA operation,
we scanned the proton beam at the target
and compared the result with that from the 250~kA operation.
Figure~\ref{fig:horn_discrepancy} shows 
the correlation between the profile center at the silicon array 
and the proton beam position at the target.
As seen in the figure, the correlation is negative at the 250~kA operation.
Whereas it
becomes positive at the 205~kA operation.
The reason is considered as follows.
An off-center proton beam produces secondary particles asymmetrically with respect 
to the beam-axis because of different path lengths
through the target.
In the case of 0~kA horn current setting,
particles in the opposite direction of the off-center beam 
are more attenuated in the target (see left in Fig.~\ref{fig:horn_focus}).
The muon beam 
would then
be directed in the same direction as the off-center beam,
resulting in the positive correlation.
On the other hand, when the horn currents
are tuned on and the focusing becomes stronger,
this results in 
the negative correlation between the profile center at 
the muon monitor 
and the proton beam position at the target.
This is considered as follows.
When the proton beam is off-center at the target, 
differences 
arise in the exit points from the target 
between secondary particles.
For example, if the proton beam hits the target in the positive direction,
the secondary particles generated 
exit in the positive direction 
from the target faster than ones generated in 
the negative direction.
Those which exit 
the target faster 
experience a larger Lorentz force
and are therefore
more focused (see the right in Fig.~\ref{fig:horn_focus}).
Thus, the muon beam would be directed in the opposite direction,
i.e. negative direction.
The MC simulation was also used to 
confirm the
dependence of the profile center 
position
of the muon beam on the proton beam position at the target 
with different horn currents.
The result also showed the correlation is positive at the 205~kA operation.
In addition, 
correlation is lost for some horn current value between 205~kA and 250~kA.
This means that the profile center of the muon beam
is no 
longer
sensitive to the proton beam position at the target
for some horn current value
between 205~kA and 250~kA.


\section{Absolute muon yield  measurement by emulsion detector\label{sec:emulsion}}

The 
particles arriving
at the muon monitor are expected to be a mixture of muons and some 
lower
energy components, namely electrons and gammas
as shown in Table~\ref{tab:pid_at_mumon}.
Since the standard detectors of the muon monitor, the silicon detectors and the ionization chambers, are designed to obtain the profile of the muon beam by measuring the integrated ionization in their 
active volumes,
the measured profile is a convolution of all components in Table~\ref{tab:pid_at_mumon}.
However, the understanding of the muon flux is a key issue to understand the nature of the beam.
Therefore, it is important to measure the absolute muon flux
and compare it with the model.
In order to complement the 
muon monitor
measurement and to diagnose the absolute muon flux, a set of emulsion detectors is temporarily inserted during the period of commissioning.

The emulsion detector has a high spatial resolution,
 down to tens of nano-meters, 
 allowing a 5D reconstruction of particle trajectories, 3 positions (X, Z, Y) and 2 angles (tan$\theta_x$, tan$\theta_y$), 
for particle densities of up to 10$^6$ particles/cm$^2$. Furthermore, by employing a proper detector structure, it can successfully reject the low energy components
by their
 multiple Coulomb scattering inside the detector materials.

The emulsion film used for this measurement 
is
the recent standard emulsion film, so-called OPERA film \cite{operafilm}, which has two sensitive 44~$\mu$m emulsion layers on both sides of a plastic base 
(205 $\mu$m thick) 
and the thickness of the film in terms of radiation length is 0.003 X$_0$.
In order to reduce the background tracks accumulated in the emulsion film, a {\it refreshing} treatment \cite{operafilm} 
was previously applied.
All films 
were put in a climate chamber and 
 remained at
 T=28${}^\circ\mathrm{C}$ 
 with
 R.H.=98\% 
 for 6 days, and then dried 
 at
 T=20${}^\circ\mathrm{C}$ 
 with
 R.H.=50\% 
 for 1 day.
The emulsion detector module consists of 8 emulsion films shaped into 6 cm $\times$ 5 cm dimensions
 and the horizontal array of 7 modules (25 cm spacing) 
was placed 
 on the neutrino beam-axis
 just downstream of the ionization chambers 
 to measure the absolute muon yield.

In addition to the above detector array, another detector dedicated 
to measuring the momentum distribution of muons 
was also placed at the center of neutrino beam. 
  These muon momentum measurements
 will be the subject of future publication.

The data readout of emulsion films is performed with the OPERA scanning microscopes~\cite{microscope} 
and the tracks crossing several films are reconstructed by the  FEDRA emulsion data analysis framework~\cite{fedra}.
The reconstructed tracks which have at least 4 hits out of 8 films, with the most upstream hit existing among the first 4 films are selected.

The performance of the detector module for the flux measurement is also checked with 
a Geant4-based MC simulation (G4). 
The flux input, described
in Sec.~\ref{sec:mc},
is propagated through the detector by G4 with the detection efficiency described later. 

\begin{figure}[tb]
  \centering
  \begin{tabular}{cc}
    \includegraphics[height=5cm]{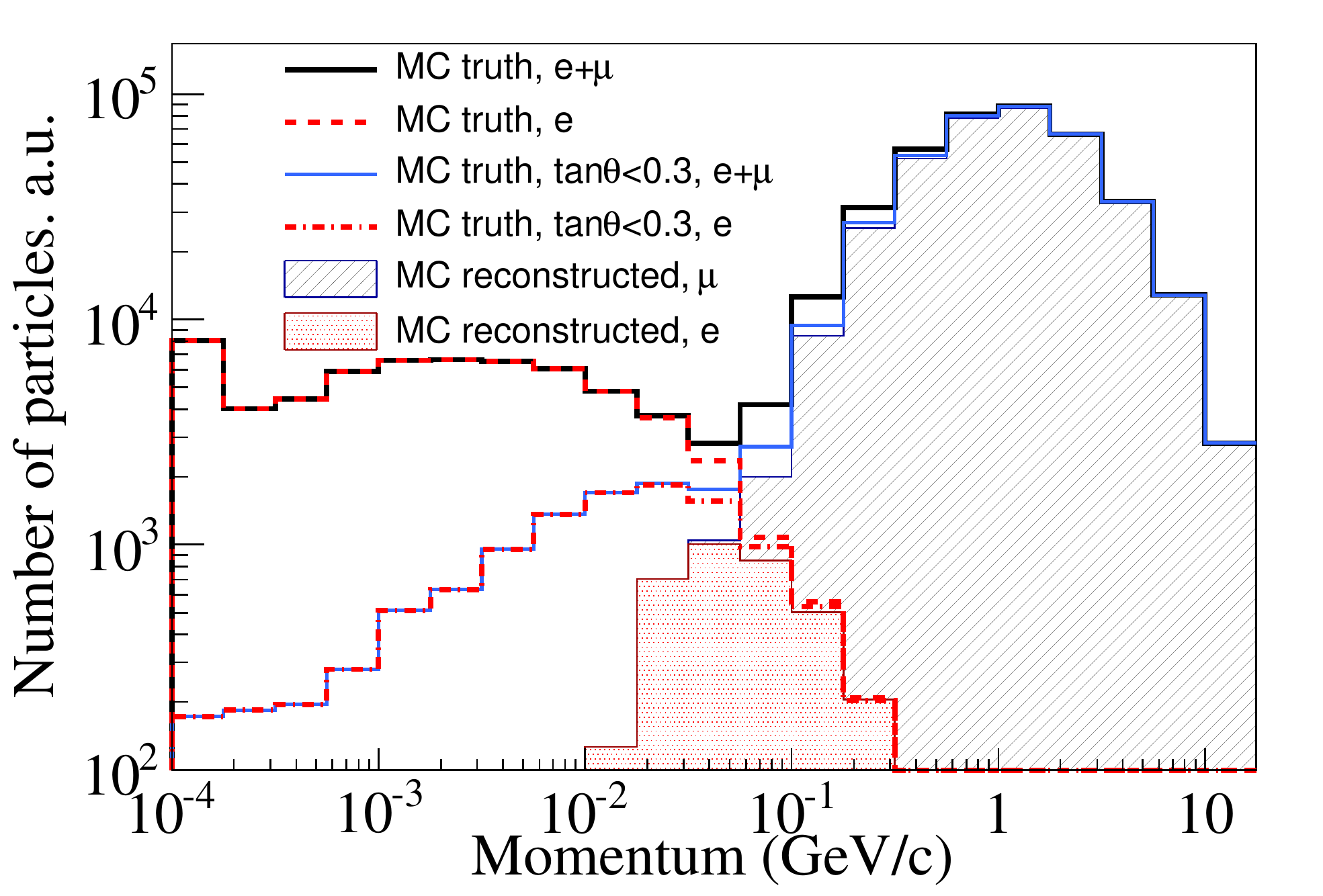} &
    \includegraphics[height=5cm]{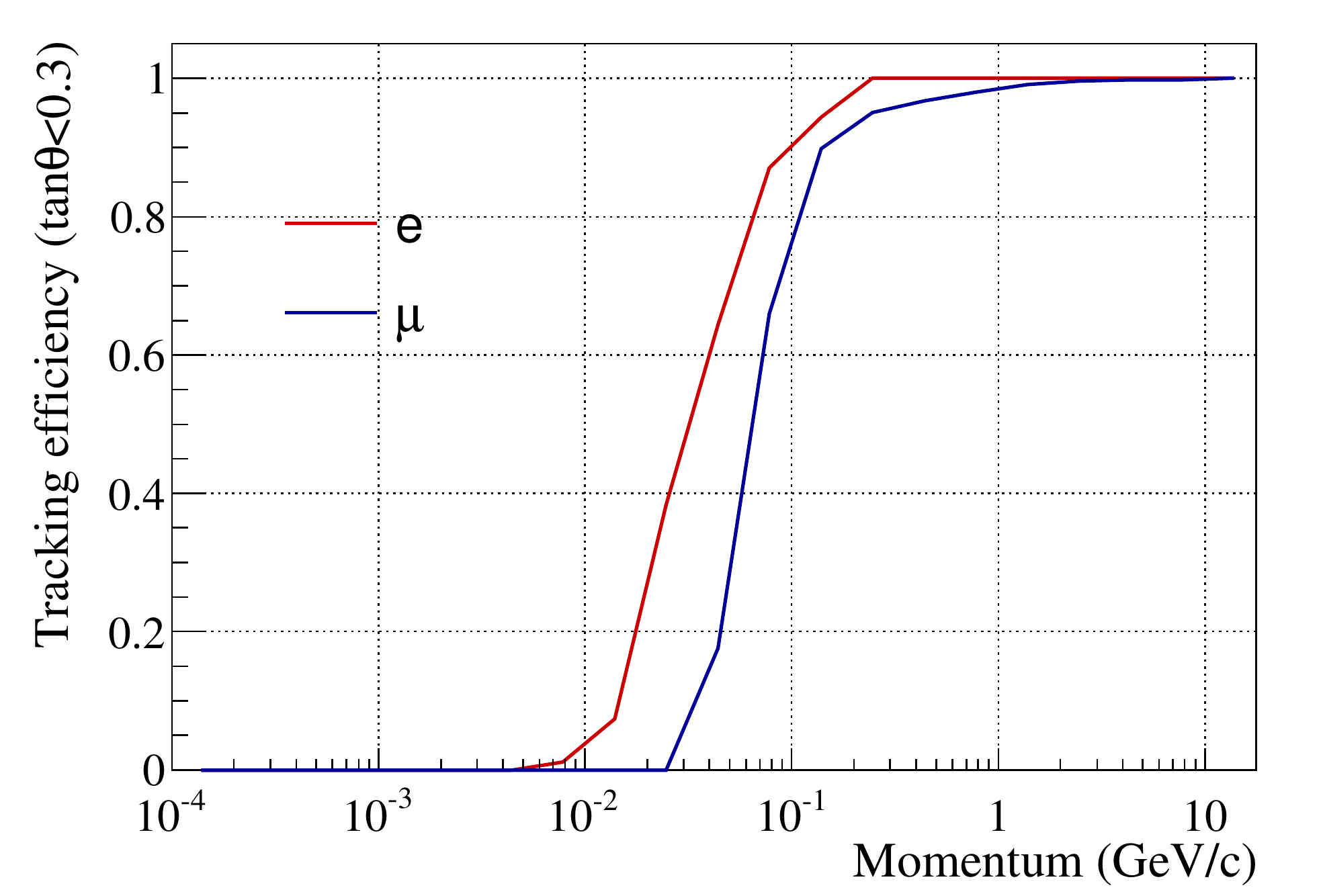} \\
  \end{tabular}
  \caption{Left: Momentum distribution of the input fluxes and the reconstructed particles. Right: tracking efficiency for the particles in an angular acceptance of $\tan\theta<0.3$.}
  \label{em:performance}
\end{figure}

The energy distributions of the input fluxes and the reconstructed particles are shown
in Fig.~\ref{em:performance} on the left,
as a stacked histogram with 
black and dashed red lines.
An application of the angular acceptance of $\tan\theta <0.3$ 
(where $\theta$ is the angle from the normal vector of the film surface) 
can effectively reduce the low energy components 
because the track angles of low energy components have less correlation 
with the beam angle
(blue and fine dashed red lines).
The reconstructed tracks are shown as a filled stacked histogram; 
the additional reduction of low energy components is achieved via their multiple Coulomb scattering in the 8 films by requesting a stringent angular matching between the films.
The track reconstruction efficiencies for muons and electrons are given in Fig.~\ref{em:performance} on the right,
 as a function of their momenta.
The overall detection efficiency for muons is estimated to be 98.0\% with respect to 
the muons in the angular acceptance or 94.2\% for muons in all angular space.
The contamination by electrons is expected to be as small as 1.0\% with respect to the number of muons
reconstructed in the angular acceptance.

The emulsion detectors 
were
exposed to a low intensity beam twice with the different horn current settings, 
see Table~\ref{em:table:exposures}. The films 
were
then photo-developed.
\begin{table}
\centering
\caption{Horn current, the number of shots,
and p.o.t. for each exposure time.}
\begin{tabular}{cccc}
\hline\hline
Exposure & Horn current & \# of shots& pot \\
\hline
A& 250 kA & 1 & 1.949$\times 10^{11}$\\
\hline
B& 0 kA   & 2 & 1.984 + 1.951 = 3.935$\times 10^{11}$\\
\hline\hline
\end{tabular}
\label{em:table:exposures}
\end{table}

\begin{figure}[htbp]
	\centering
	\begin{tabular}{cc}
	\includegraphics[height=5cm]{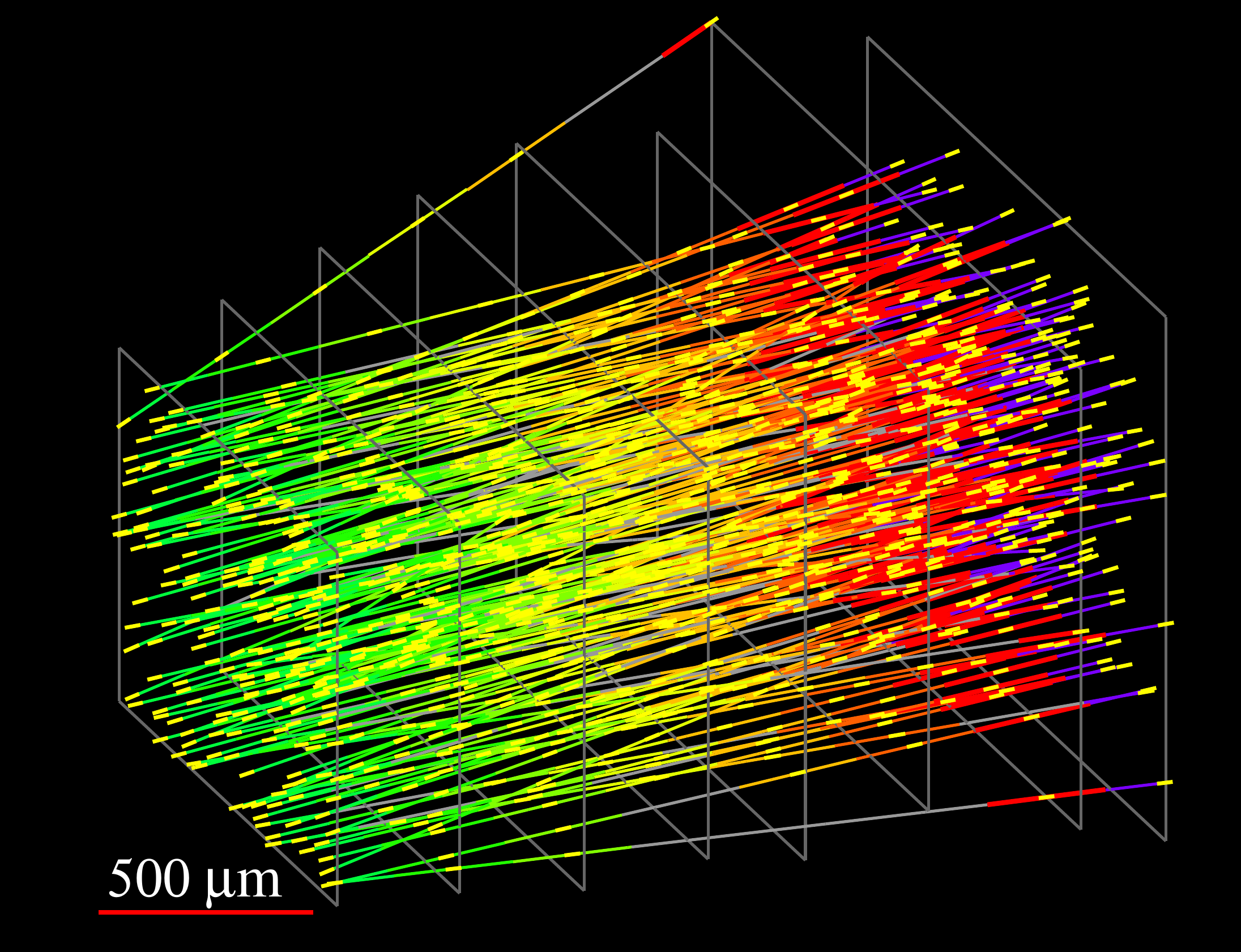} & 
	\includegraphics[height=5cm]{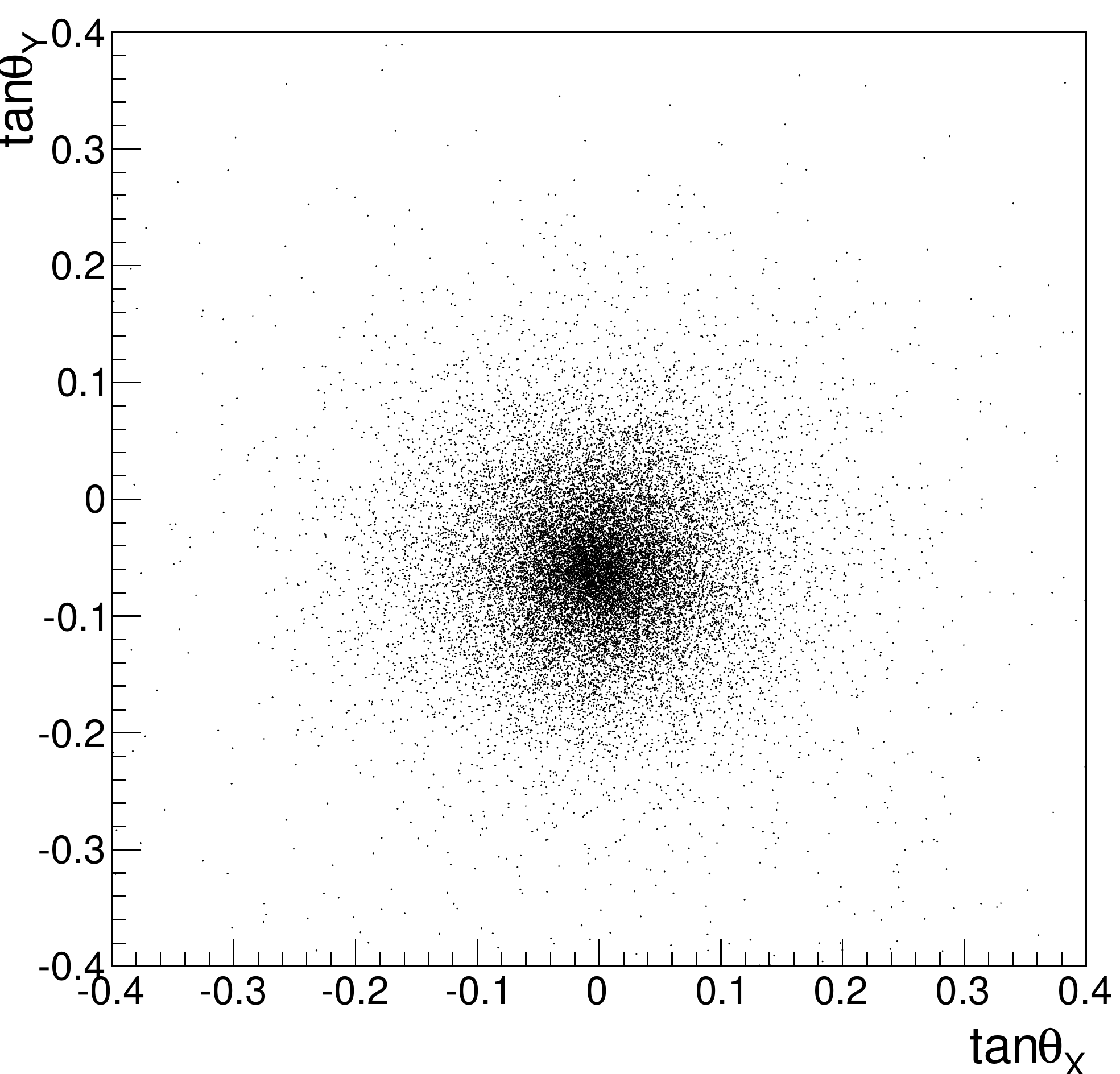} \\
	\end{tabular}
	\caption{Left: Example of reconstructed tracks entering in the 1 $\times$ 1 mm$^2$ surface in the center module when the horn is operated at 250~kA. The color of lines shows the depth in the module. Right: The measured angular distribution in the same detector. Each dot corresponds to the individual track angle.}
	\label{em:reconstructed}
	\end{figure}

For each film, the data is taken 
from a
2 cm$^2$ area at the center of film. 
The relative alignments between the films are found by using the beam tracks themselves with sub-micron precision. 
After the track reconstruction, an effective area of 1 cm$^2$ at the center of film is used to compute the flux.
An example of the reconstructed tracks is shown in Fig.~\ref{em:reconstructed} on the left, and the angular distribution on the right.

\begin{figure}[htbp]
	\centering
	\includegraphics[width=1.\textwidth]{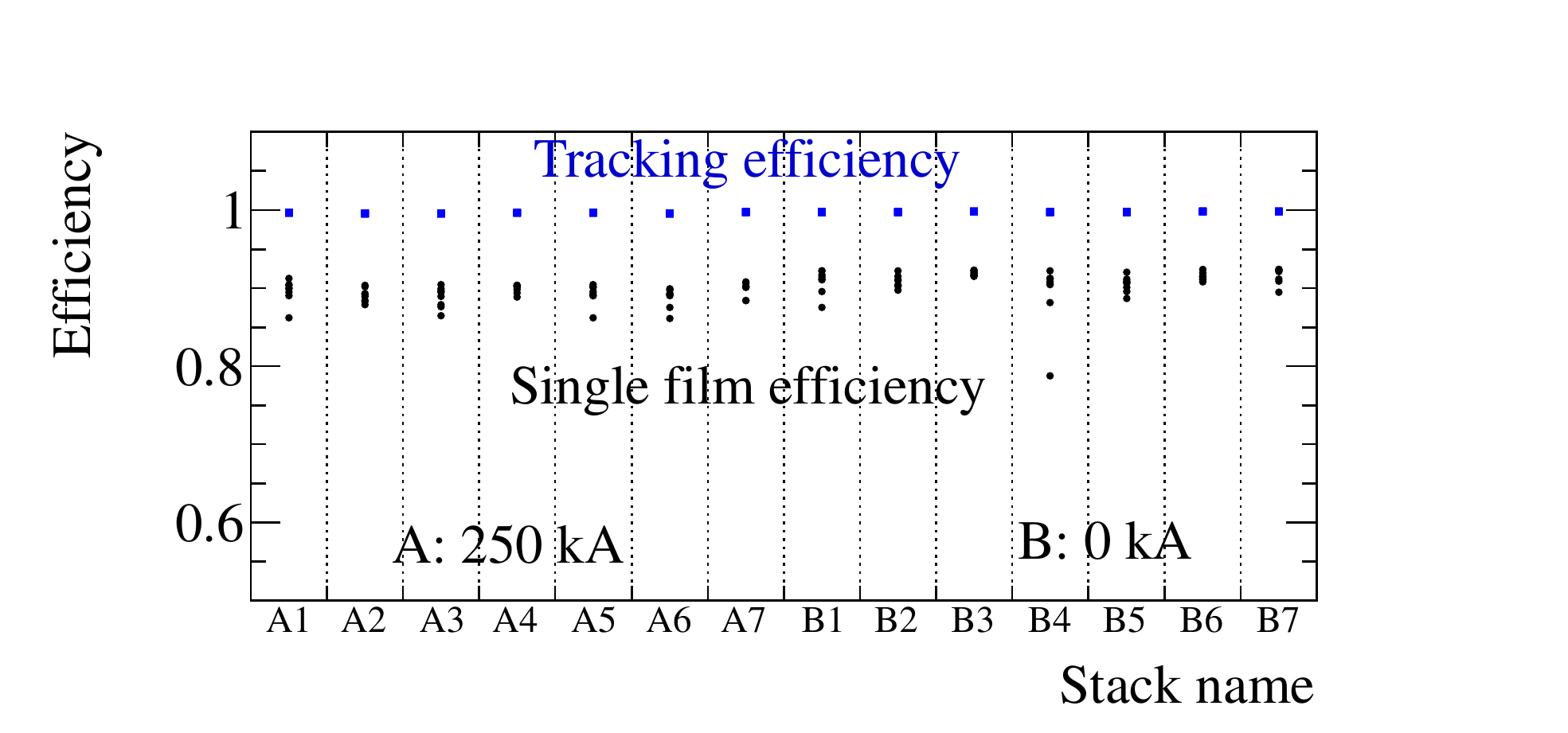}
	\caption{Single film and tracking efficiencies for all films and modules. The efficiency is computed using all tracks with the angular acceptance of $\tan\theta < 0.3$.}
	\label{em:eff}
	\end{figure}

The detection efficiencies of 
each film and module are measured 
using the reconstructed tracks 
in the module, counting the number of missing hits 
in the film for the tracks crossing the film and are shown in Fig.~\ref{em:eff}.
The tracking efficiency of each module is then computed 
by taking account of the efficiencies of individual films in the module.
 The average track detection efficiencies for high energy particles, 
 where multiple Coulomb scattering does not play a role for the inefficiency, are calculated to be higher than 99.5\% for all modules. The flux data is corrected by the track detection efficiency module-by-module in the later analyses. 

\begin{figure}[htbp]
	\centering
	\includegraphics[width=0.8\textwidth]{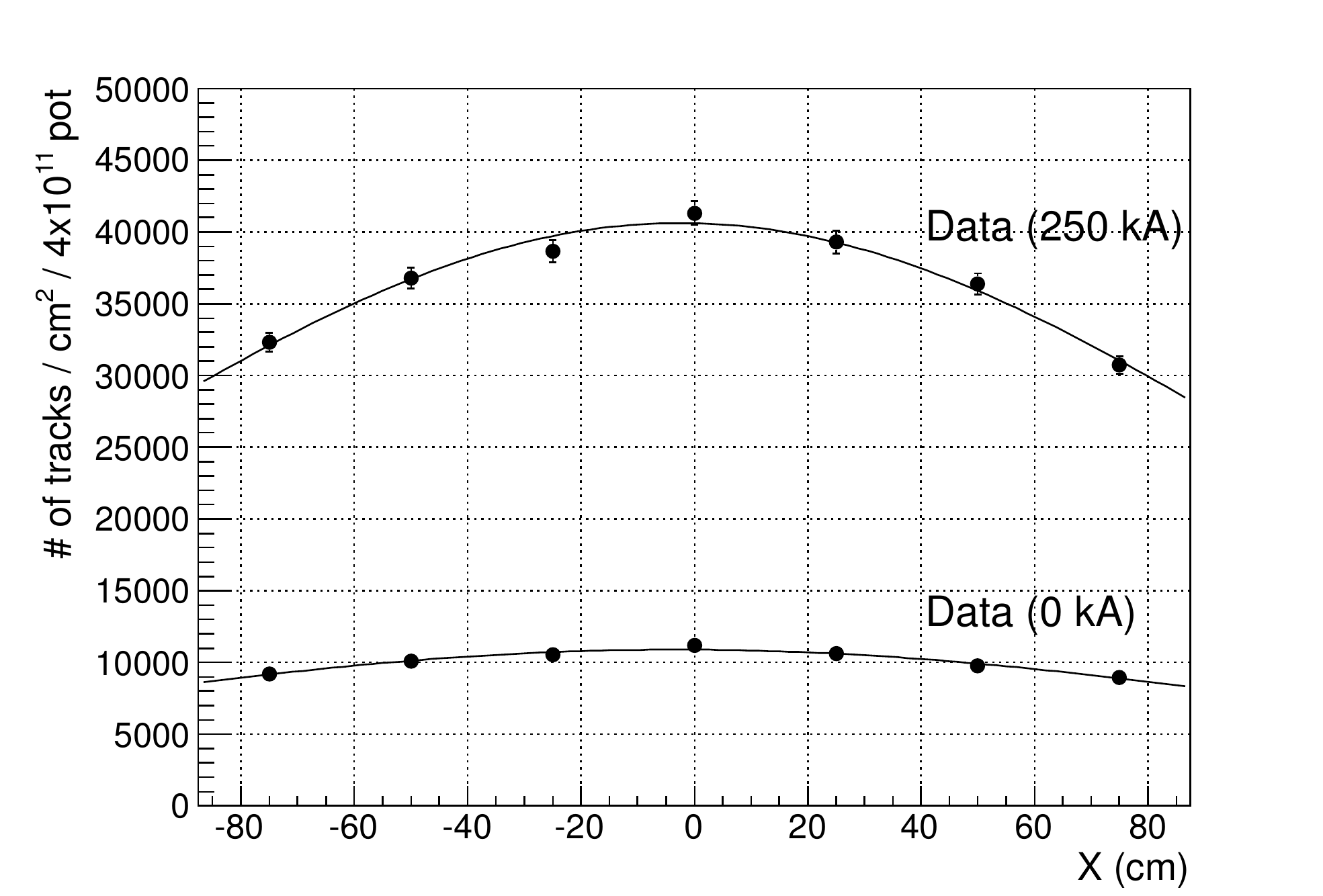}
	\caption{ Measured muon flux
	with fitted lines 
	represented as solid curves.
	Error bars denote statistical and systematic ones.}
	\label{em:dataprof}
	\end{figure}

The measured 
muon flux normalized to $4\times 10^{11}$ p.o.t. is shown in Fig.~\ref{em:dataprof}. 
For each data point, the systematic uncertainty of 2\%,
which were estimated from
the measurement reproducibility test\footnote{Two flux modules were placed one after the other and exposed to the beam. 
The difference of number of muons between those modules was assigned to the systematic error.}, 
is taken into account in addition to the statistical error. 
The flux data is fitted with a Gaussian function and the measured muon fluxes at the profile center are 
($1.09\pm 0.01)\times 10^{4}$ tracks/cm$^{2}/4\times10^{11}$~p.o.t. when the horns are not operated and increased to ($4.06\pm 0.05)\times 10^{4}$ tracks/cm$^{2}/4\times 10^{11}$~p.o.t. when the horns are operated at 250~kA. 
The 1$\sigma$ widths of the 
flux profiles are measured to be 122.4$\pm$ 6.5~cm
 and 105.6 $\pm$ 4.1~cm, respectively.

%
%
%
%
%

\section{Comparison of the muon yield with 
prediction based on tuned-simulation\label{sec:mc}}

\begin{figure}[tb] 
   \centering
   \includegraphics[width=0.6\textwidth]{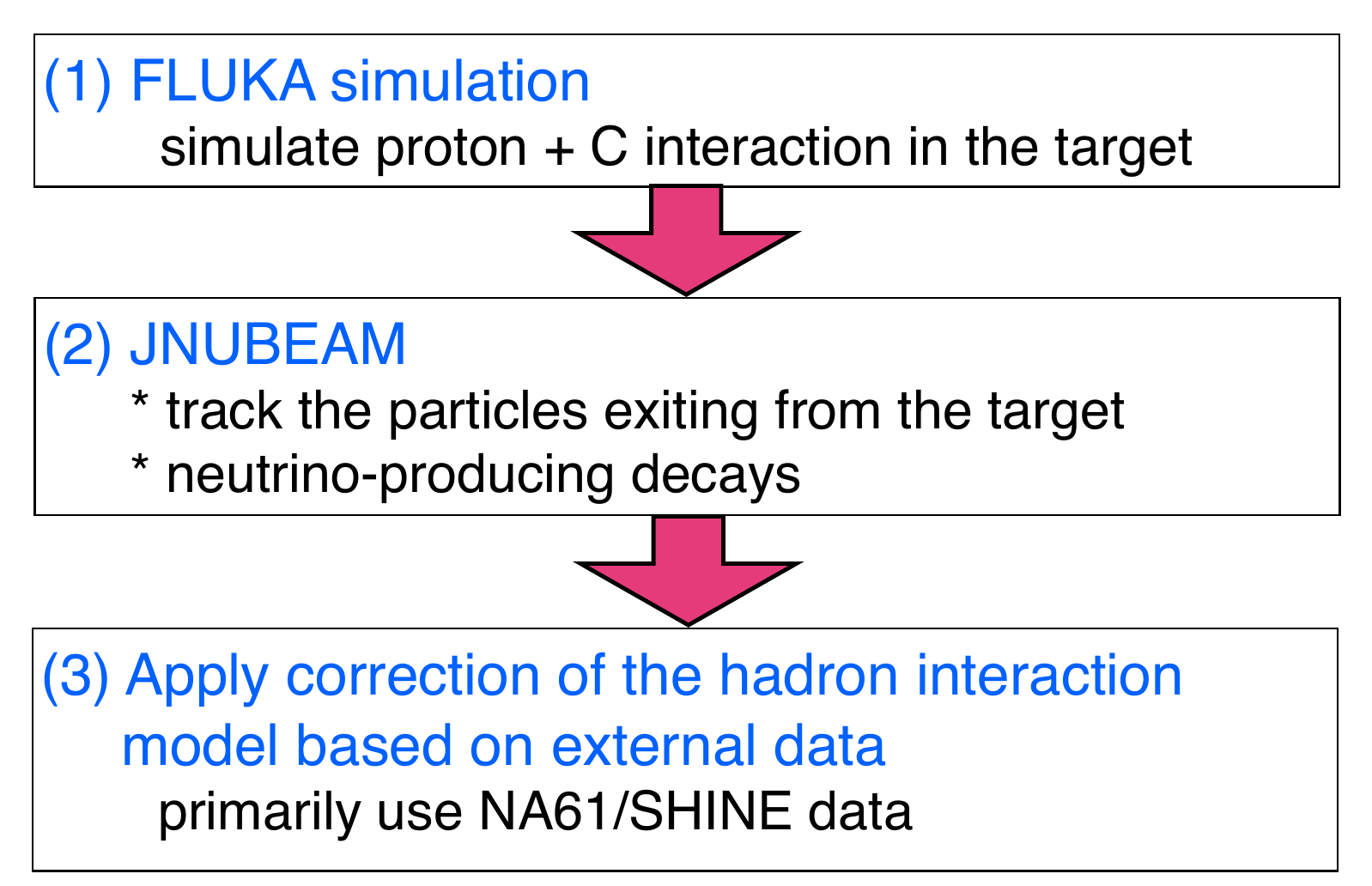} 
   \caption{Flow diagram of the flux prediction.}
   \label{fig:flux_flow}
\end{figure}

As described in Sec.~\ref{subsec:mc_prep},
T2K uses FLUKA2008 for the simulation of the hadronic interaction in the 
graphite target
and the kinematic information for 
the particles 
is then transferred to the JNUBEAM simulation.
Hadronic interactions in the JNUBEAM simulation are treated with GCALOR.
For a precise prediction of the neutrino and muon flux,
T2K
corrects the model based on
hadron interaction data provided by external experiments,
primarily relying on the NA61/SHINE measurements~\cite{na61exp_rev}.
A flow diagram for the precise estimation of the muon flux 
is shown in Fig.~\ref{fig:flux_flow}.
This section first describes how the muon flux is 
predicted
in Sec.~\ref{subsec:tune_muflux}.
Systematic errors 
of the prediction
summarized in Sec.~\ref{subsec:systerr_muflux}.
The result is then compared with the measurement from the emulsion data
in Sec.~\ref{subsec:mc_emulsion_comparison}.

\subsection{Correction of the muon flux\label{subsec:tune_muflux}}

In order to make a prediction based on the external hadron-interaction data,
we use the method developed for the T2K flux prediction~\cite{flux}.
Here, we briefly summarize the procedure.
Following quantities modeled in FLUKA2008 and GCALOR are corrected based on 
the external data,
\begin{enumerate}
\item
interaction rates for $p$, $\pi^{\pm}$ and $K^{\pm}$, and
\item
differential production of $\pi^{\pm}$, $K^{\pm}$ and $K^0_{L}$
in the interaction of protons on the target.
\end{enumerate}

The NA61/SHINE measurement provides
both the differential production
and the interaction rate~\cite{Abgrall:2011ae,PhysRevC.85.035210},
which are primarily used for the prediction of the neutrino and muon flux.
Other experimental data are used to 
compensate 
for the measurement of NA61/SHINE~\cite{eichten,allaby,e910}.

The hadronic interaction rate is defined as 
a cross section
calculated by subtracting 
the cross section for
the quasi-elastic scattering process
($\sigma_\text{qe}$)
from the total inelastic cross section ($\sigma_\text{inel}$):
\begin{equation}
\sigma_\text{prod} = \sigma_\text{inel} - \sigma_\text{qe}
\end{equation}

Most of the
data provides $\sigma_\text{inel}$.
Thus, $\sigma_\text{qe}$ 
are
subtracted from $\sigma_\text{inel}$
to extract $\sigma_\text{prod}$.
Since the prediction of FLUKA2008 was found to be good agreement with the data,
this 
correction
is applied only to $\sigma_\text{prod}$ in GCALOR.

\begin{figure}[tb] 
   \centering
   \includegraphics[width=0.48\textwidth]{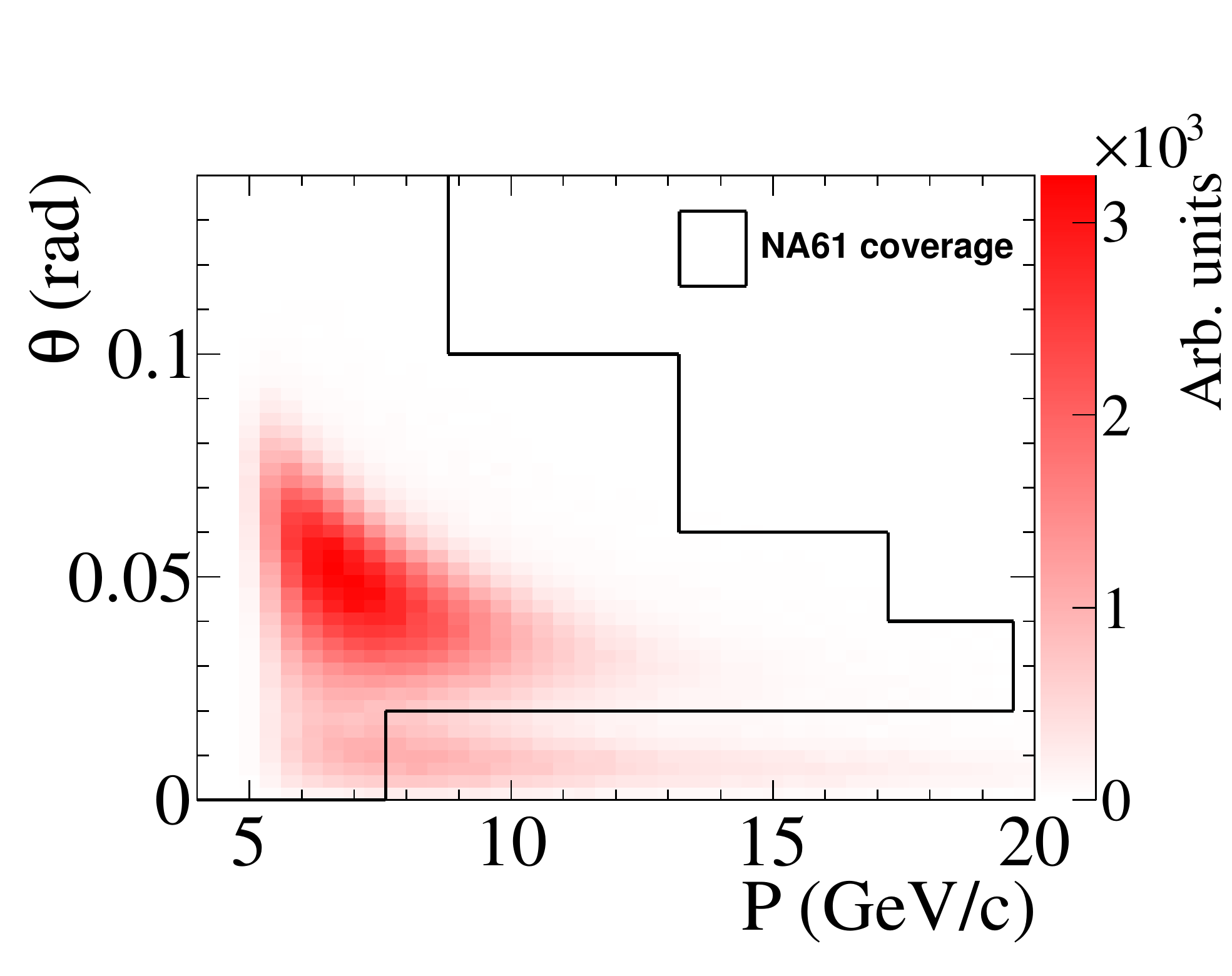} 
   \includegraphics[width=0.48\textwidth]{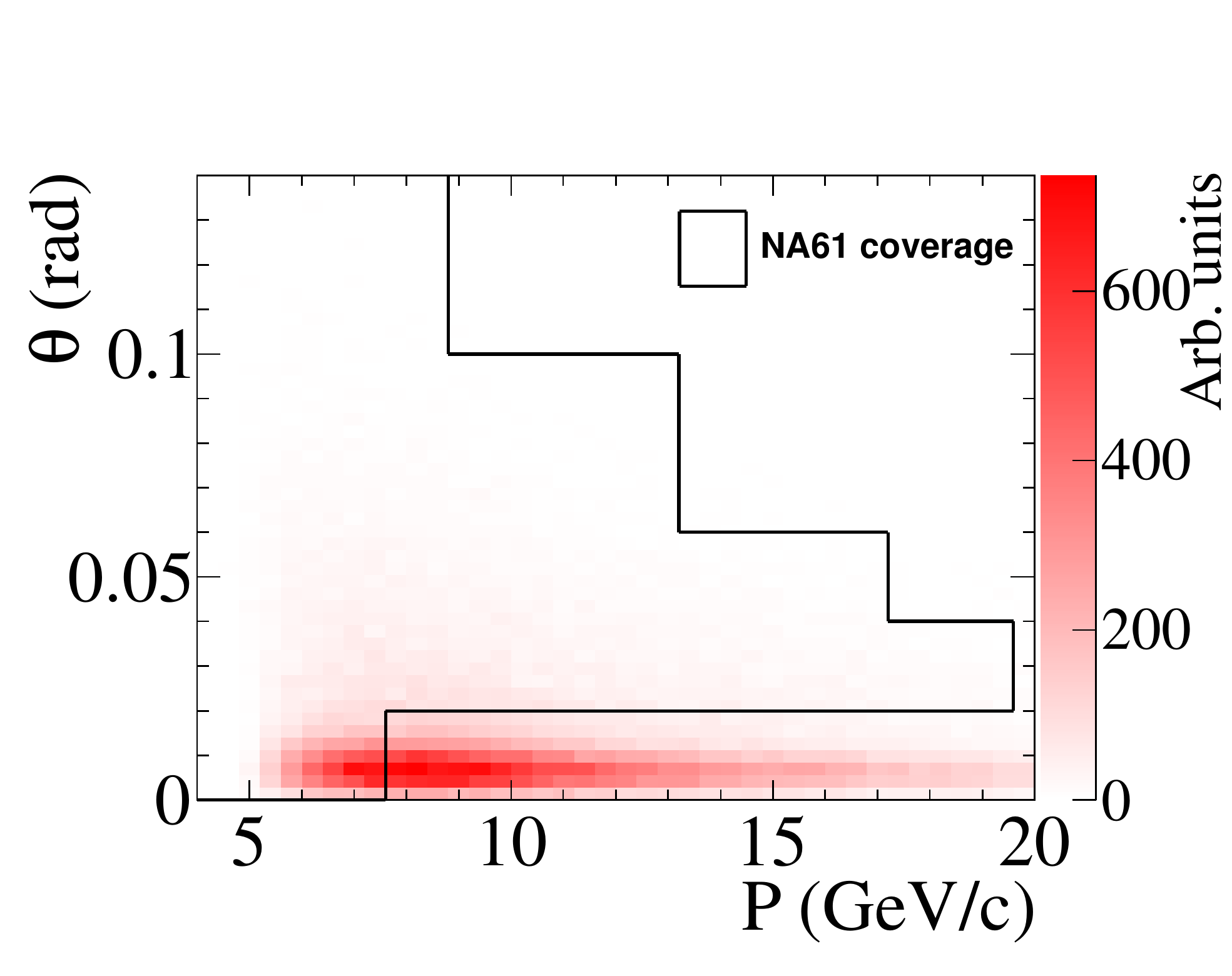} 
   \caption{Phase space of the parent $\pi^{+}$s contributing muon flux at the muon monitor
   when the horn currents are operated at 250~kA (left) and 0~kA (right).
   NA61 coverages shown in these figures correspond to the 2007 data.}
   \label{fig:mumon_ptheta}
\end{figure}

Figure~\ref{fig:mumon_ptheta} shows the phase space 
of the parent $\pi^{+}$ contributing to the muon flux at the muon monitor
when the horn currents are 
set at 250~kA (left) and 0~kA (right).
Most of the phase space 
is covered by the NA61/SHINE data for the 250~kA operation.
On the other hand,
only $\pi^{+}$s 
in the forward angle regions 
reach the muon monitor for the 0~kA operation.
This results in 
only around 30\% coverage by the NA61/SHINE data.
Most of the $K^+$ contributing to the muon flux 
are not covered by the NA61/SHINE data~\footnote{The correction of the flux was performed using results from the NA61/SHINE measurement in 2007. NA61/SHINE also collected data in 2009, where statistics increased by an order of magnitude as compared to the 2007 data and a phase space coverage was enlarged. Therefore the flux is expected to be predicted more precisely with the 2009 data.}.
The differential production
depends on the incident particle momentum, $p_{in}$, 
and target nucleus, $A$.
For secondary $\pi^{\pm}$s produced by 31~GeV/c protons in the phase space 
covered by NA61/SHINE data,
corrections are directly applied using the NA61/SHINE data.
The corrections for tertiary pion production from secondary particles
and for the production at materials ($A$) other than graphite are obtained 
with extrapolations from the NA61/SHINE data 
assuming momentum and A-dependent scaling~\cite{Feynman69,bmpt_paper,barton,skubic}.


The correction for the production of $K^{+}$ and $K^{-}$
in the phase space not covered by the NA61/SHINE data
is estimated with other experimental data~\cite{eichten,allaby}.
For the hadrons in phase space uncovered by any experimental data,
the corrections are no longer applied.

\begin{figure}[tb]
  \centering
  \includegraphics[width=0.9\textwidth]{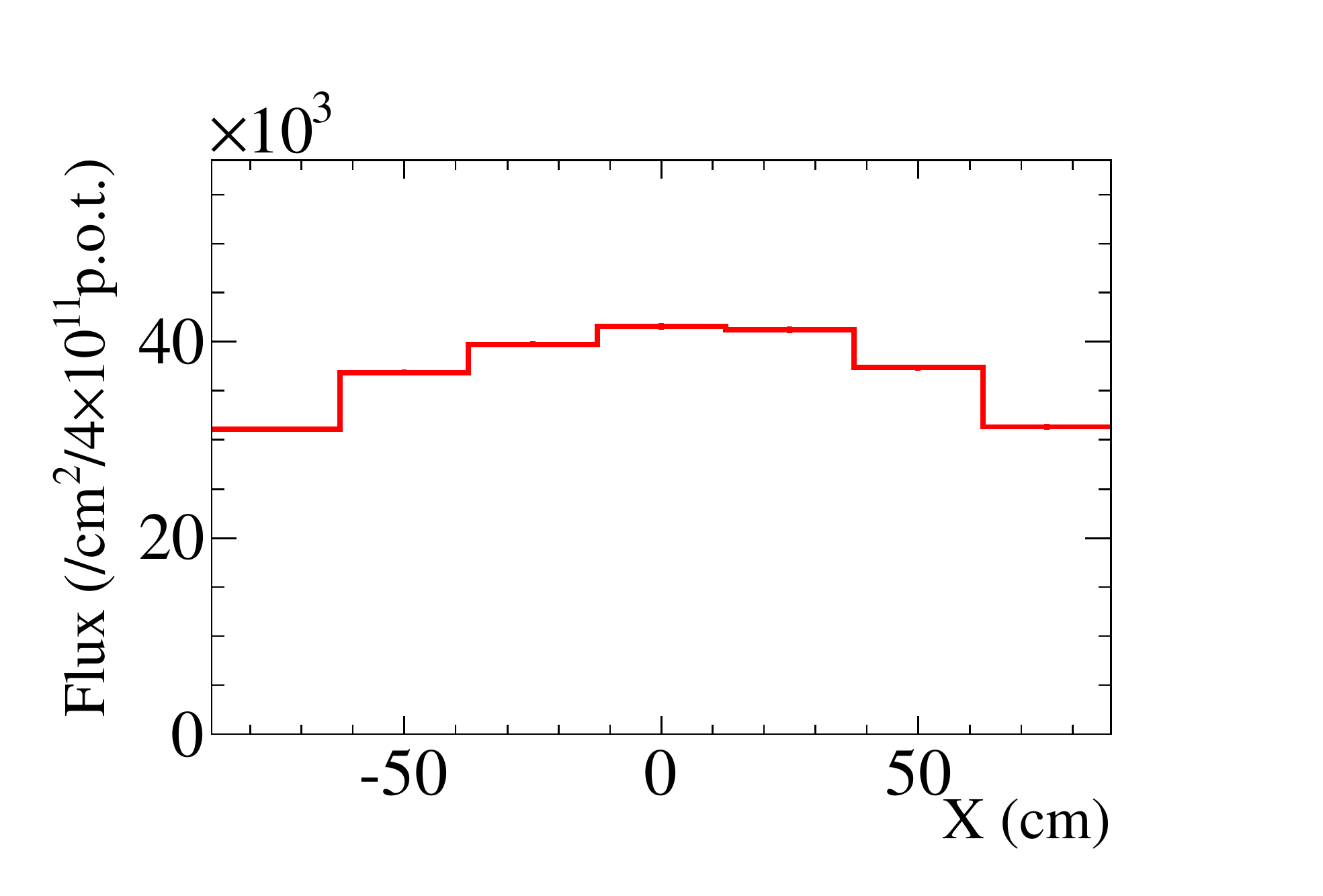} 
   \caption{Predicted muon flux at the emulsion for the 250~kA horn operation.}
   \label{fig:muprof_tuned}
\end{figure}

Figure~\ref{fig:muprof_tuned} and Table~\ref{tab:muprof_tuned}
show the absolute muon flux at the emulsion 
after the 
correction
of the hadron productions.
As a result of the correction
the absolute muon flux is increased by about 20\%
(1.9\% by the production cross section, 
14.8\% by the pion and 3.1\% by the kaon production tuning).

\begin{table}[tb]
\centering
\caption{Predicted 
muon flux at the emulsion 
with 250~kA horn current.
The second and third rows show the fitted peak and sigma
of the profile of the muon beam.}
\begin{tabular}{ll}
\hline \hline
& Flux after correction\\
\hline
 Muon flux (/cm$^2$/10$^{12}$p.o.t.) & $9.72\times10^4$\\
 Fitted peak of the profile (/cm$^2$/10$^{12}$p.o.t.) & $1.09\times10^5$\\
Fitted sigma of the profile (cm) & $100.6\pm1.8$\\
\hline\hline
\end{tabular}
\label{tab:muprof_tuned}
\end{table}

%
%

\subsection{Systematic error on the flux prediction\label{subsec:systerr_muflux}}

The systematic error 
on the muon flux prediction originates from
uncertainty in
the hadron production
and measurement error of the proton beam,
horn current,
and alignment of the target.

\subsubsection{Uncertainty of the hadron production}

The systematic error on the production cross section
is dominated by the uncertainty of the quasi-elastic subtraction.
This assumption is based on 
discrepancies in the production data
among data sets~\cite{flux}.

The systematic error of the pion or kaon differential production
comes from:
\begin{enumerate}
\item measurement error of the pion/kaon differential production,
\item uncertainty in the momentum or target scaling,
\item uncertainty from the pion/kaon production in the phase space not covered by data.
\end{enumerate}

In addition to uncertainties listed above, 
the systematic error on the muon flux also arises from uncertainty in secondary nucleon productions.
The error is basically estimated using other experimental data sets~\cite{eichten,allaby}.
For the region where the incident protons undergo 
with a small momentum transfer,
we assign 100\% error to the production 
due to the lack of relevant data.

The muon flux is also generated from the interactions
at the beam dump (C) and it contributes about 4\% of the total flux as shown in Table~\ref{tab:muon_parents_material}.
We test correction of
the proton-beam production at the dump
and it results in decreasing the muon flux by 0.7\%.
The change of 0.7\% is assigned to the systematic error on the muon flux.

The muon flux is also generated from decays of $\Lambda$, $\Sigma$ or other particles
whose productions are not corrected.
According to the MC simulation, 0.6\% of the muons come from 
such particles.
Since there is no relevant data for 
such production, we conservatively assign a 100\% error
on the production. 
In addition, 0.6\% of muons come from decays 
of quaternary particles, which are not corrected at this stage
because of the small contributions to the muon flux.
A 100\% error is conservatively assigned to the decay mode.
Finally we take account of 1.2\% for the systematic errors from these productions.

In total, 
we attribute 13.4\% of the systematic error 
in the muon flux
to uncertainty in
the hadron production 
and summarize those errors in Table~\ref{tab:haderr}.

\begin{table}[tb]
\centering
\caption{Systematic errors on the muon flux due to uncertainty in the hadron production} 
\begin{tabular}{ll}
\hline\hline
Error source & Error size \\
\hline
Pion production & 9.0\% \\
Kaon production & 1.3\% \\
Production cross section & 9.1\% \\
Secondary nucleon production & 3.6\% \\
Dump interaction & 0.7\% \\
Decays from $\Lambda$, $\Sigma$, and quaternary particles & 1.2\%\\
\hline
Total & 13.4\% \\
\hline\hline
\end{tabular}
\label{tab:haderr}
\end{table}

\subsubsection{Uncertainty of the proton beam measurement}
A trajectory and optics of the proton beam
are measured by the proton-beam monitors placed in the primary beam line as described in Sec.~\ref{sec:t2kexperiment}.
In the MC simulation, the parameters of the proton beam are varied within those errors
attributing to the measurement errors from the proton beam monitors.
The resultant variation of the muon flux (0.9\%) is then estimated at the muon monitor
and is taken as the systematic error on the muon flux.
In addition, 2.6\% error in the p.o.t. measurment, which mainly originates from 
the calibration accuracy of the beam monitor, is assigned to the error in the muon flux.
In total, we take account of 2.8\% for the systematic error on the muon flux.

\subsubsection{Uncertainty of the absolute horn current}
During beam operation, the monitored values of the horn current were found to be
drifted by 2\%, 5~kA.
the drift is considered to be mainly due to the temperature dependence in the hardware monitoring.
In the MC simulation, the horn currents are varied by 5~kA from the nominal values (=250~kA).
The variation of the muon flux, 3.6\%, is then taken as the systematic error on the muon flux.

\subsubsection{Uncertainty of the target alignment}
The rotation of the target with respect to the horn-axis was surveyed 
and was measured to be 1.3~mrad (0.1~mrad) in the horizontal (vertical) direction.
The effect of the target alignment is estimated by rotating the target in the simulation
according to the measured values described above.
The resultant variation of the muon flux, 0.5\%, is assigned to the systematic error.

\subsubsection{Skin effect in the magnetic horns}

Since the horn current is applied as pulses of about 1~ms,
the current would flow only around the surface of the conductor
due to the skin effect.
However, the 
present
MC simulation assumes a flat current density.
To estimate the size of
the skin effect, 
the magnetic field in the simulation is modified by taking the skin depth into account.
The modification results in decreasing the muon flux by 2.0\%.
The change is assigned to the additional systematic error.

\subsubsection{Summary of the systematic error on the absolute muon flux}

Table~\ref{tab:flux_syst_summary}
summarizes the systematic error 
on the absolute muon flux measurement.
Finally we assigned a total of 14.3\% error to the absolute muon flux.
In the case of 0~kA horn current setting,
the systematic error cannot be fully evaluated because the phase space of pions (kaons)
is poorly covered by data.
This may result in a large error size for the muon flux.
From these reasons, we evaluated the systematic error only for the 250~kA operation.

\begin{table}[tb]
\centering
\caption{Summary of the systematic error on the absolute muon flux.}
\begin{tabular}{l l}
\hline
\hline
Error source & Error size \\
\hline
Hadron production & 13.4\% \\
Proton beam & 2.8\% \\
Absolute horn current & 3.6\% \\
Target alignment & 0.5\% \\
Horn skin effect & 2.0\% \\
MC statistics & 0.3\% \\
\hline
Total & 14.3\% \\
\hline
\hline
\end{tabular}
\label{tab:flux_syst_summary}
\end{table}

\begin{figure}[tb]
  \centering
   \includegraphics[width=0.8\textwidth]{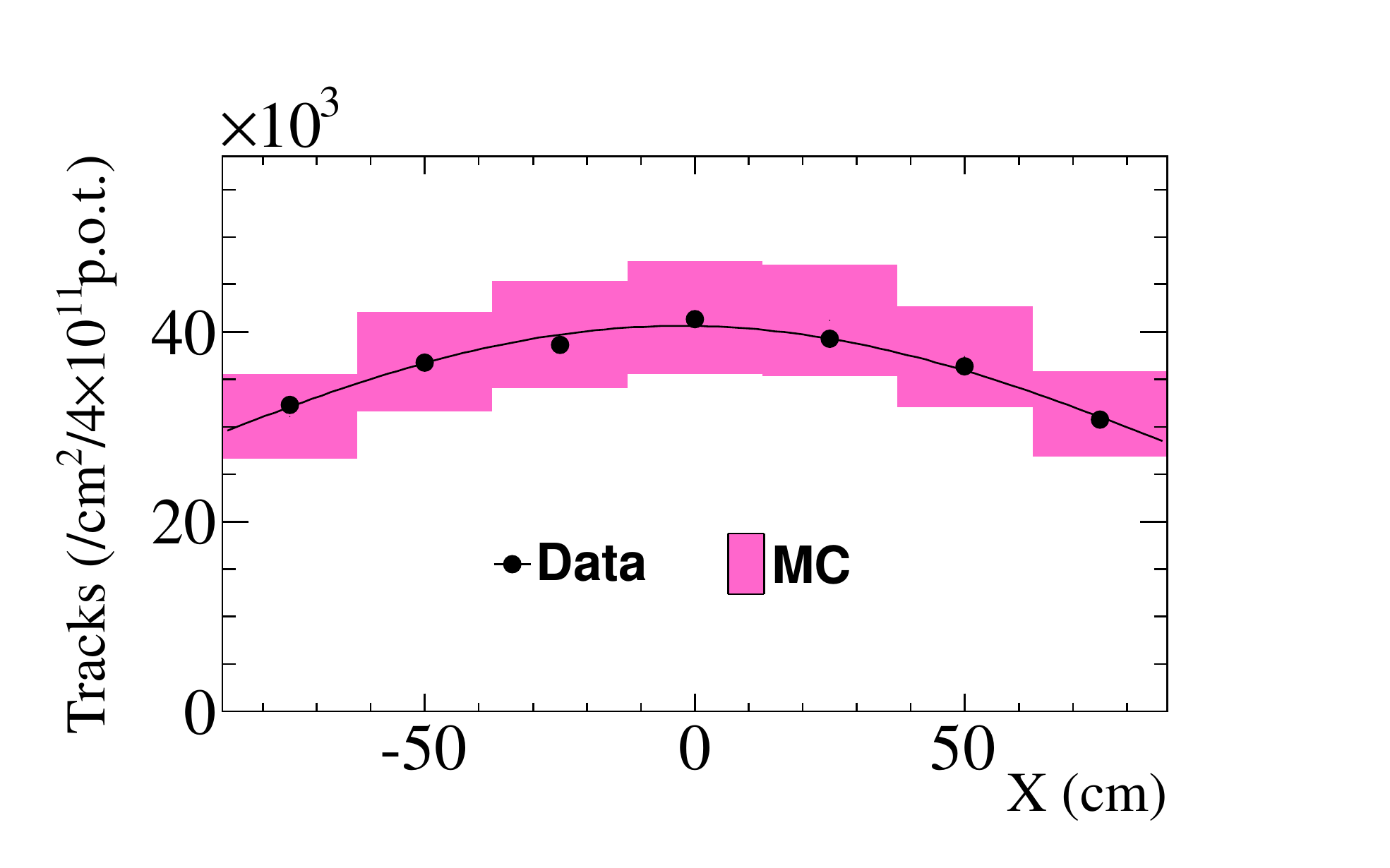} 
   \caption{Comparison in the muon flux at the emulsion
between the measurement and prediction
for the 250~kA horn operation.
The band shows the uncertainty of the prediction.
}
   \label{fig:mc_emulsion_comparison}
\end{figure}

\subsection{Comparison with the emulsion measurement\label{subsec:mc_emulsion_comparison}}

Figure~\ref{fig:mc_emulsion_comparison} 
shows the comparison 
of the reconstructed profile of the muon flux at the emulsion
between the measurement and 
prediction
where the hadron production is 
corrected
as described in Sec.~\ref{subsec:tune_muflux}.
Both of the profiles are obtained at 250~kA horn operation.
Table~\ref{tab:mc_emulsion_comparison} 
summarizes
comparisons in the muon flux at the emulsion detectors
between the data and 
prediction.
In the case of the 250~kA operation,
the ratio of the 
measured muon flux 
to the prediction is
$0.971\pm0.143$.
The data and prediction agrees quite well.
In case of the 0~kA operation,
there is about 
a 10\% discrepancy in the
flux between the measurement and prediction.
This is because the phase space of secondary pions contributing 
to the muon flux at the emulsion is less
constrained by the external data.
This result demonstrates that 
our understanding of the muon, and hence neutrino, production 
is quite good for the 250~kA horn setting.

\begin{table}[t]
\centering
\caption{Comparison of the muon flux at the emulsion detector
between the measurement and prediction. The fluxes are normalized to $4\times10^{11}$~p.o.t.}
\scalebox{0.83}
{
\begin{tabular}{l | ccc | ccc}
\hline\hline
& \multicolumn{3}{c|}{0~kA} & \multicolumn{3}{c}{250~kA} \\
\cline{2-7}
& flux  & flux ratio  & fitted profile & flux  & flux ratio  & fitted profile \\
& tracks /cm$^2$ 
& (Data/MC) &  width (cm)  & tracks /cm$^2$ & (Data/MC) &  width (cm) \\
\hline
Data     
& $10892 \pm 126 $ & - & $122.4 \pm 6.5$ 
& $40628 \pm 468 $ & - & $105.6 \pm 4.1$ \\
T2K MC
& 9682  & 1.12 & 100.3  
& 41833  & 0.971 & 98.7\\
\hline
\hline\hline
\end{tabular}
}
\label{tab:mc_emulsion_comparison}
\end{table}



\section{Conclusion\label{sec:conclusion}}

In this paper, we described the measurement of the beam direction by the muon monitor
and the study of the muon yields with the emulsion detector.
The systematic error of the beam direction measurement with the muon monitor was estimated 
to be 0.28~mrad
and fulfills our requirement
of $<0.3$~mrad.
The muon beam direction has been stable, and controlled within 0.3~mrad for most of the span of beam operation 
and measured with good resolutions.
As a consequence, we have controlled the neutrino beam direction
well within 1.0~mrad and provided good quality beam data 
to be used as
 an input to the neutrino oscillation measurements.
The muon monitor has also played in an important role in surveying the configuration of the beamline components.
To confirm our understanding of muon beam and neutrino beam,
the absolute muon flux was measured with the emulsion detector with a precision of 2\%.
It was then compared with 
prediction based on external hadron interaction data.
As a result, we obtained good agreement between the data and prediction.
This result confirms the validity of the beam control by the muon monitor
  and also demonstrates the validity of the T2K neutrino flux
  tuning based on the external hadron production data.

\section*{Acknowledgment}

We thank the J-PARC staff for superb accelerator performance and the
CERN NA61 collaboration for providing valuable particle production data.
We also would like to thank the J-PARC neutrino beam group for their putting 
great efforts to keep taking high quality beam data
and the Nagoya University for helping us with the emulsion detector construction.
We acknowledge the support of MEXT, Japan, and 
SNSF and Canton of Bern, Switzerland.
Special thanks go to Nikhul Patel, who carefully reads this manuscript.
In addition, authors have been further supported by funds from JSPS, Japan.

\bibliographystyle{ptephy} 
\bibliography{mumon}

\begin{thebibliography}{10}

\bibitem{t2k}
K.~Abe et~al. (T2K~collaboration), Nucl.\ Instrum.\ Meth., {\bf A659}, 106--135
  (2011).

\bibitem{mumon}
K.~Matsuoka et~al., Nucl.\ Instrum.\ Meth., {\bf A624}, 591--600,
  doi:10.1016/j.nima.2010.09.074 (2010).

\bibitem{HORN}
A.K. Ichikawa, Nucl.\ Instrum.\ Meth., {\bf A690}, 27--33 (2012).

\bibitem{ingrid}
K.~Abe et~al. (T2K~collaboration), Nucl.\ Instrum.\ Meth., {\bf A694}, 211--223
  (2012).

\bibitem{jesse}
S.J. Harris and C.E. Doust, Radiat.\ Res., {\bf 66}, 11 (1976).

\bibitem{Ziock}
H.J. Ziock et~al., IEEE Trans.\ Nucl.\ Sci.\ NS-40, 344 (1993).

\bibitem{Higuchi:2003xs}
T.~Higuchi et~al., eConf C {\bf 0303241}, TUGT004, hep-ex/0305088 (2003).

\bibitem{Ferrari:2005zk}
A.~Ferrari et~al., CERN-2005-010 (2005).

\bibitem{flux}
K.~Abe et~al., Phys.\ Rev.\ D, {\bf 80}, 012001 (2013).

\bibitem{GEANT3}
R.~Brun, F.~Carminati, and S.~Giani, CERN-W5013 (1994).

\bibitem{GCALOR}
C.~Zeitnitz and T.~A. Gabriel, In Proc. of International Conference on
  Calorimetry in High Energy Physics (1993).

\bibitem{otr}
M.~Hartz et~al., Nucl.\ Instrum.\ Meth., {\bf A703}, 45--58 (2013).

\bibitem{operafilm}
T.~Nakamura et~al., Nucl.\ Instrum.\ Meth., {\bf A556}, 80--86 (2006).

\bibitem{microscope}
N.~Armenise et~al., Nucl.\ Instrum.\ Meth., {\bf A551}, 261--270 (2005).

\bibitem{fedra}
V.~Tyukov et~al., Nucl.\ Instrum.\ Meth., {\bf A559}, 103--105 (2006).

\bibitem{na61exp_rev}
N.~Abgrall et~al. (NA61/SHINE~Collaboration), JINST, {\bf 9}, P06005 (2014).

\bibitem{Abgrall:2011ae}
N.~Abgrall et~al. (NA61/SHINE~Collaboration), Phys.\ Rev.\ C, {\bf 84}, 034604
  (2011).

\bibitem{PhysRevC.85.035210}
N.~Abgrall et~al. (NA61/SHINE~Collaboration), Phys.\ Rev.\ C, {\bf 85}, 035210
  (2012).

\bibitem{eichten}
T.~Eichten et~al., Nucl.\ Phys.\ B, {\bf 44}, 333 (1972).

\bibitem{allaby}
J.~V. Allaby et~al.,
\newblock High-energy particle spectra from proton interactions at 19.2
  {GeV/c},
\newblock Technical Report 70-12, CERN (1970).

\bibitem{e910}
I.~Chemakin et~al., Phys.\ Rev.\ C, {\bf 77}, 015209 (2008).

\bibitem{Feynman69}
R.P. Feynman, Phys.\ Rev.\ Lett., {\bf 23}, 1415 (1969).

\bibitem{bmpt_paper}
M.~Bonesini et~al., Eur.\ Phys.\ J.\ C, {\bf 20}, 13 (2001).

\bibitem{barton}
D.~S. Barton et~al., Phys.\ Rev.\ D, {\bf 27}, 2580 (1983).

\bibitem{skubic}
P.~Skubic et~al., Phys.\ Rev.\ D, {\bf 18}, 3115 (1978).

\end{thebibliography}

\end{document}